\documentclass[11pt,preprintnumbers,nofootinbib,amsmath,amssymb]{revtex4}
\usepackage{float}
\usepackage{indentfirst}
\usepackage{amsmath,amssymb,epsfig,subfigure}
\usepackage{graphicx}
\usepackage[export]{adjustbox}
\usepackage{color}
\usepackage{multirow}
\usepackage{booktabs}
\usepackage{changes}
\usepackage{footnote}
\usepackage{mathrsfs}
\usepackage[shortcuts]{extdash}

\usepackage{hyperref}
\hypersetup{colorlinks=true,linkcolor=blue,citecolor=magenta}

\begin{document}
\renewcommand{\baselinestretch}{1.15}
\makeatletter
\renewcommand{\thesubfigure}{\alph{subfigure}}% (a) -> a
\renewcommand{\@thesubfigure}{(\thesubfigure)\hskip\subfiglabelskip}% a -> a)
\makeatother

\title{Tidal Love numbers of multi-state Boson stars}

\preprint{}

\author{Xin-Lei Zhao, Jun-Ru Chen, and Yong-Qiang Wang\footnote{E-mail: yqwang@lzu.edu.cn, corresponding author}}

\affiliation{$^{1}$ Lanzhou Center for Theoretical Physics, Key Laboratory of Theoretical Physics of Gansu Province, School of Physical Science and Technology, Lanzhou University, Lanzhou 730000, China\\
 $^{2}$ Institute of Theoretical Physics $\&$ Research Center of Gravitation, Lanzhou University, Lanzhou 730000, China}

%\pacs{bla}

\begin{abstract}
In this paper, we calculate the tidal Love numbers of multi-state boson stars (MSBSs) composed of ground state and first excited state complex scalar fields. Under synchronized and nonsynchronized frequency conditions, the background solutions of MSBSs are classified into single-branch and double-branch types. The field functions, ADM mass, and binding energy of different solutions are discussed. We then calculate the quadrupolar ($\ell=2$) electric and magnetic tidal Love numbers for branches containing stable solutions. Our results show that the electric tidal Love numbers are initially positive and then suddenly transition to negative values. This phenomenon occurs when the parameters satisfy $\tilde{\mu}_1 > 0.891$ or $\tilde{\omega}_0 > 0.777$; for smaller values of these parameters, the electric Love numbers remain positive. The magnetic tidal Love numbers are always negative, with absolute values smaller than those of the electric tidal Love numbers.
 
\end{abstract}

%\keywords

\maketitle
\newpage
%\tableofcontents

\section{INTRODUCTION}\label{Sec1}

Tidal effects play an essential role in the evolution of binary systems. During the inspiral phase, as gravitational wave emission carries away orbital energy, the orbital separation decreases and tidal interactions become significant. These interactions induce tidal deformations of the compact objects, quantified by the tidal Love numbers, and produce observable imprints on the gravitational waveform~\cite{Mora:2003wt,DeLuca:2025bph}. The Love numbers extracted from gravitational wave signals can reveal information about the internal structure, such as the neutron star equation of state~\cite{Flanagan:2007ix}.

The tidal Love numbers were originally introduced to describe Earth's deformation under external tides~\cite{love1909yielding}. Two dimensionless parameters, $h$ and $k$, were defined to describe the vertical deformation and the gravitational potential deformation, respectively. Subsequently, T. Shida introduced another parameter, $l$, to describe the lateral deformation~\cite{shida1912body}. These Love numbers are applicable within Newtonian gravity. In 2008, the quadrupolar ($\ell=2$) tidal Love numbers of fully relativistic neutron stars were calculated, and the tidal correction to the gravitational waveform phase was found to appear at the fifth post-Newtonian order~\cite{Hinderer:2007mb,Flanagan:2007ix}. In the relativistic theory, the Love numbers $k$ are classified by parity into electric and magnetic types, defined respectively by the ratios of the induced mass and current multipole moments to the multipole moments of the tidal field~\cite{Damour:2009vw}. Interestingly, subsequent studies showed that Schwarzschild black holes have vanishing tidal Love numbers~\cite{Poisson:2005pi,Fang:2005qq,Binnington:2009bb}, and this result also holds for slowly rotating black holes~\cite{Pani:2015hfa,Landry:2015zfa,Gurlebeck:2015xpa,Poisson:2014gka}. This is because black holes lack internal structure or elastic deformation mechanisms to respond to tidal fields. The vanishing tidal Love numbers of black holes to some extent distinguish them from other compact objects~\cite{Cardoso:2017cfl,Sennett:2017etc,Cardoso:2019rvt}.

Unlike black holes, boson stars are another type of exotic compact object. The study of boson stars dates back to the 1960s. D. J. Kaup found that a Klein-Gordon geon, obtained by coupling a complex scalar field to Einstein gravity, could exist and avoid gravitational collapse~\cite{Kaup:1968zz}. Around the same time, R. Ruffini and S. Bonazzola obtained similar solutions by coupling a real quantized scalar field to gravity~\cite{Ruffini:1969qy}. These solutions later became known as boson stars. In 2015, R. Brito et al. studied Proca stars formed by coupling a Proca field to gravity~\cite{Brito:2015pxa}. Boson stars have been extensively studied, including models such as self-interacting boson stars~\cite{Schunck:1999zu,Kling:2017hjm,Sanchis-Gual:2021phr,Brito:2024biy}, charged boson stars~\cite{Jetzer:1989av,Jetzer:1989us,Jetzer:1992tog}, rotating boson stars~\cite{Yoshida:1997qf,Kleihaus:2005me,Gervalle:2022fze}, and oscillating boson stars~\cite{Seidel:1991zh,Yang:2025yej}. Research on boson stars has provided new insights into compact objects and cosmology; for instance, boson stars can mimic black holes~\cite{Guzman:2009zz,Barranco:2011wq,Herdeiro:2021lwl,Bambi:2025wjx} and are considered candidates for dark matter~\cite{Lee:1995af,Suarez:2013iw,Chen:2020cef,Mourelle:2024dlt}. Studies show that the tidal Love numbers of boson stars are smaller than those of neutron stars~\cite{Cardoso:2017cfl}. The tidal Love numbers of Proca stars differ from those of boson stars, and this difference is imprinted in the gravitational wave signals from binary mergers~\cite{Herdeiro:2020kba}. Furthermore, the Love numbers of other compact objects, such as gravastars~\cite{Pani:2015tga,Uchikata:2016qku}, axion stars~\cite{Chen:2023vet}, regular black holes~\cite{Coviello:2025pla}, and fermion soliton stars~\cite{Berti:2024moe}, have also been studied. The calculation of tidal Love numbers can also be extended to rotating compact objects~\cite{Pani:2015hfa,Landry:2015zfa,Landry:2017piv,Chia:2020yla} and modified gravity theories~\cite{Yazadjiev:2018xxk,Meng:2021ijp,Yang:2022ees,Cano:2025zyk}. Other studies of tidal Love numbers include the I-Love-Q relations~\cite{Yagi:2013bca,Yagi:2013awa}, as well as dynamical tidal Love numbers~\cite{Hinderer:2016eia,Steinhoff:2016rfi,Perry:2023wmm,HegadeKR:2024agt} and nonlinear tidal Love numbers~\cite{DeLuca:2023mio,Pani:2025qxs}. 

Most studies to date have focused primarily on compact objects composed of a single field. However, multi-field compact objects are also of considerable interest. In 2010, A. Bernal et al. constructed multi-state boson stars (MSBSs) composed of two complex scalar fields, respectively in the ground state and the first excited state~\cite{Bernal:2009zy}. It is generally believed that single excited state boson stars are unstable and will decay into the ground state or a black hole under perturbations~\cite{Balakrishna:1997ej}. Studies on MSBSs show that mixing the ground state and the first excited state can yield stable solutions. Unlike Ref.~\cite{Bernal:2009zy}, we study MSBSs in both synchronized and nonsynchronized frequency cases, investigate the properties of the solutions, and calculate the binding energies to study stability. Then, we focus on the tidal deformability of MSBSs that contain stable solutions and calculate the tidal Love numbers.

This paper is organized as follows. In Sec. \ref{Sec2}, we construct the MSBSs by minimally coupling Einstein gravity to two complex scalar fields. Sec. \ref{Sec3} introduces the even-parity and odd-parity perturbations to the background of the MSBSs and obtains the corresponding perturbation equations. The background solutions are presented in Sec. \ref{Sec4}, and the numerical results for the quadrupolar tidal Love numbers of MSBSs are shown in Sec. \ref{Sec5}. Finally, Sec. \ref{sec6} contains our summary and discussion. We adopt the metric signature $(-,+,+,+)$ and use natural geometric units with $c=G=1$.

\section{THE MODEL OF MULTI-STATE BOSON STARS}\label{Sec2}

We consider a system consisting of two complex scalar fields minimally coupled to Einstein gravity. The action for this system is given by~\cite{Bernal:2009zy}
 \begin{equation}\label{equS}
     S=\int \sqrt{-g} \, d^{4} x\left(\frac{R}{16 \pi }+\mathcal{L}_{0}+\mathcal{L}_{1}\right),     
\end{equation}	
where $R$ is the Ricci scalar. $\mathcal{L}_{0}$ and $\mathcal{L}_{1}$ are the Lagrangians for the ground and first excited state fields:
\begin{equation}\label{Lag}
	\mathcal{L}_{0}=-g^{\alpha \beta} \bar{\Phi}_{0, \alpha} \Phi_{0, \beta}-\mu_0^2 \bar{\Phi}_{0} {\Phi}_{0}\, ,  
    \quad\mathcal{L}_{1}=-g^{\alpha \beta} \bar{\Phi}_{1, \alpha} \Phi_{1, \beta}-\mu_1^2 \bar{\Phi}_{1} {\Phi}_{1}\, , 
\end{equation}
where $\Phi_{0}$ and $\Phi_{1}$ are the complex scalar fields for the ground and first excited states, respectively.

Varying the action (\ref{equS}) with respect to the metric and the field functions yields the Einstein field equations and the matter field equations, which are given as follows:
\begin{equation}\label{equE}
    R_{\alpha \beta} - \frac{1}{2}g_{\alpha \beta}R = 8\pi (T^{0}_{\alpha \beta} + T^{1}_{\alpha \beta})\, ,
\end{equation}
\begin{equation}\label{equM}
	\Box \Phi_{n} - \mu_n^2 \Phi_{n}=0\,, \quad n=0,1\,, 
\end{equation}
where $T^{0}_{\alpha \beta}$ and $T^{1}_{\alpha \beta}$ are the energy-momentum tensors of the scalar fields in the ground state and the first excited state,
\begin{equation}\label{equT0}
	T_{\alpha \beta}^{0}=\bar{\Phi}_{0, \alpha} \Phi_{0, \beta}+\bar{\Phi}_{0, \beta} \Phi_{0, \alpha}-
	g_{\alpha \beta}\left[\frac{1}{2} g^{\gamma \delta}\left(\bar{\Phi}_{0, \gamma} \Phi_{0, \delta}+
	\bar{\Phi}_{0, \delta} \Phi_{0, \gamma}\right)+\mu_0^2 \bar{\Phi}_{0} {\Phi}_{0}\right]\,, 
\end{equation}
\begin{equation}\label{equT1}
	T_{\alpha \beta}^{1}=\bar{\Phi}_{1, \alpha} \Phi_{1, \beta}+\bar{\Phi}_{1, \beta} \Phi_{1, \alpha}-
	g_{\alpha \beta}\left[\frac{1}{2} g^{\gamma \delta}\left(\bar{\Phi}_{1, \gamma} \Phi_{1, \delta}+
	\bar{\Phi}_{1, \delta} \Phi_{1, \gamma}\right)+\mu_1^2 \bar{\Phi}_{1} {\Phi}_{1}\right]\,.
\end{equation}
The action (\ref{equS}) remains invariant under a global $U(1)$ transformation ${\Phi}_{n}\rightarrow e^{i\alpha}{\Phi}_{n}$, where $\alpha$ is a constant. Therefore, according to Noether's theorem, there exist conserved currents in the system,
\begin{equation}\label{equJ}
	J_n^{\mu} = -i\left(\bar{\Phi}_{n}\partial^\mu\Phi_{n} - \Phi_{n}\partial^\mu\bar{\Phi}_{n}\right)\,, \quad n=0,1\,.
\end{equation}
By integrating the timelike component of these conserved currents over a spacelike hypersurface $\varSigma$, we obtain the conserved Noether charge,
\begin{equation}\label{equQ}
	Q_n = \int_{\varSigma}J_n^t\,, \quad n=0, 1\,.
\end{equation}

For a static, spherically symmetric background, the metric is assumed to be~\cite{Macedo:2013jja}
\begin{equation}\label{metric}
   ds^2 = -e^{u(r)} dt^2 + e^{v(r)} dr^2 + r^2 (d\theta^2 + \sin^2\theta \, d\phi^2)\,,
\end{equation}
and the functions $u(r)$ and $v(r)$ depend only on the radial coordinate $r$, since the metric is static. The ansatz for the scalar field is taken as~\cite{Liebling:2012fv}
\begin{equation}\label{equf}
	\Phi_n^{(0)} = \phi_{n}(r)e^{-i\omega_nt}\,, \quad n=0,1\,,
\end{equation}
where $n$ is the number of radial nodes of the field function. Here, we only consider $n=0$ and $n=1$, corresponding to the ground state and the first excited state, with frequencies $\omega_0$ and $\omega_1$, respectively.

Substituting the above metric and field ansatz into the Eqs. (\ref{equE}) and (\ref{equM}), we obtain the following ordinary differential equations
\begin{equation}\label{equu}
\begin{split}
   u^{\prime}(r) = \frac{-1 + e^{v(r)}}{r} &- 8\pi re^{v(r)}\left(\mu_0^2 \phi_0(r)^2 + \mu_1^2 \phi_{1}(r)^2\right) + 8\pi r e^{-u(r)+v(r)} \left( \omega_0^2 \phi_0(r)^2 + \omega_1^2 \phi_{1}(r)^2 \right) 
   \\ & + 8\pi r \left( \phi_{0}^{\prime}(r)^2 + \phi_{1}^{\prime}(r)^2 \right)\, , 
\end{split}
\end{equation}
\begin{equation}\label{equv}
\begin{split}
   v^{\prime}(r) = \frac{1 - e^{v(r)}}{r} &+ 8\pi re^{v(r)}\left(\mu_0^2 \phi_0(r)^2 + \mu_1^2 \phi_{1}(r)^2\right) + 8\pi r e^{-u(r)+v(r)} \left( \omega_0^2 \phi_0(r)^2 + \omega_1^2 \phi_{1}(r)^2 \right) 
   \\ & + 8\pi r \left( \phi_{0}^{\prime}(r)^2 + \phi_{1}^{\prime}(r)^2 \right)\, , 
\end{split}
\end{equation}
\begin{equation}\label{equf0}
    \phi_0''(r) = -\left(-e^{v(r)} \mu_0^2 + e^{-u(r)+v(r)} \omega_0^2\right) \phi_0(r) - \left(\frac{2}{r} + \frac{u'(r)}{2} - \frac{v'(r)}{2}\right)\phi_0'(r) \, ,
\end{equation}
\begin{equation}\label{equf1}
    \phi_1''(r) = -\left(-e^{v(r)} \mu_1^2 + e^{-u(r)+v(r)} \omega_1^2\right) \phi_1(r) - \left(\frac{2}{r} + \frac{u'(r)}{2} - \frac{v'(r)}{2}\right)\phi_1'(r) \, .
\end{equation}
From Eqs. (\ref{equQ}) and (\ref{metric}), we have the Noether charge of the system as
\begin{equation}\label{equQn}
	Q_n = 8\pi\int_0^\infty e^{\frac{1}{2}(-u(r) + v(r))} r^2 \omega_n \phi_n(r)^2 dr\,, \quad n=0,1\,.
\end{equation}

\section{TIDAL PERTURBATION}\label{Sec3}

We treat the tidal perturbation of MSBSs as a first-order linear perturbation. Therefore, the perturbed metric in a spherically symmetric background can be expressed as
\begin{equation}\label{equg}
     g_{\alpha\beta} = g^{(0)}_{\alpha\beta} + h_{\alpha\beta}\,,     
\end{equation}
where $g_{\alpha\beta}^{(0)}$ is the background metric, given by Eq. (\ref{metric}). 
The perturbed metric $h_{\alpha\beta}$ can be divided into even and odd perturbations according to parity
\begin{equation}\label{equh}
      h_{\alpha\beta} = h^{(e)}_{\alpha\beta} + h^{(o)}_{\alpha\beta}\,.     
\end{equation}
Under the Regge–Wheeler gauge, the explicit forms of $h^{(e)}_{\alpha\beta}$ and $h^{(o)}_{\alpha\beta}$ are given by~\cite{Regge:1957td}
\begin{equation}\label{equhe}
h_{\alpha\beta}^{(e)} = 
\begin{pmatrix}
e^{u(r)} H_0^{\ell m}(r) & 0 & 0 & 0 \\
0 & e^{v(r)} H_2^{\ell m}(r) & 0 & 0 \\
0 & 0 & r^2 K^{\ell m}(r) & 0 \\
0 & 0 & 0 & r^2 \sin^2\theta \, K^{\ell m}(r)
\end{pmatrix}
Y^{\ell m}(\theta, \varphi)\,,
\end{equation}

\begin{equation}\label{equho}
h_{\alpha\beta}^{(o)} = 
\begin{pmatrix}
0 & 0 & h_0^{\ell m}(r) F^{\ell m}_{\theta} & h_0^{\ell m}(r) F^{\ell m}_{\varphi} \\
0 & 0 & h_1^{\ell m}(r) F^{\ell m}_{\theta} & h_1^{\ell m}(r) F^{\ell m}_{\varphi} \\
h_0^{\ell m}(r) F^{\ell m}_{\theta} & h_1^{\ell m}(r) F^{\ell m}_{\theta} & 0 & 0 \\
h_0^{\ell m}(r) F^{\ell m}_{\varphi} & h_1^{\ell m}(r) F^{\ell m}_{\varphi} & 0 & 0
\end{pmatrix}\,. 
\end{equation}
where $F^{\ell m}_{\theta} = -\frac{1}{\sin\theta} \frac{\partial Y^{\ell m}(\theta,\varphi)}{\partial \varphi}$ and $F^{\ell m}_{\varphi} = \sin\theta \frac{\partial Y^{\ell m}(\theta,\varphi)}{\partial \theta}$. Here, $Y^{\ell m}$ denotes the spherical harmonics. For convenience, we only consider the case $m=0$. 
The perturbed scalar fields are given by
 \begin{equation}\label{equfn}
     \Phi_n=\Phi^{(0)}_n+\delta\Phi_n\,, \quad n=0,1\,,     
\end{equation}
where $\Phi_n^{(0)}$ is given by Eq. (\ref{equf}). The perturbation parts of the scalar fields take the form~\cite{Mendes:2016vdr}:
\begin{equation}\label{equfp}
     \delta \Phi_n(t, r, \theta, \varphi) = \sum_{\ell,m} e^{-i\omega_n t} \frac{\psi_n(r)}{r} Y^{\ell m}(\theta, \varphi), \quad n=0,1\,.     
\end{equation}
The perturbed metric functions and scalar field functions satisfy the Einstein field equations and matter field equations under linear perturbations, which read
\begin{equation}\label{equEp}
   \delta G^{\alpha}_{\ \beta} = 8\pi \delta T^{\alpha}_{\ \beta}\,,
\end{equation}
\begin{equation}\label{equMp}
	\delta(\Box \Phi_{n} - \mu_{n}^{2} \Phi_{n})=0\,, \quad n=0,1\,,
\end{equation}

Since the background spacetime is spherically symmetric and non-rotating, the even-parity and odd-parity metric perturbations decouple. Even-parity and odd-parity perturbations induce electric-type and magnetic-type tidal deformations. These two types of deformations can be described by the relationship between the multipole moments of the tidal field and the induced multipole moments of the compact object. In order to describe the electric-type and magnetic-type tidal deformations of MSBSs more accurately and conveniently, we adopt the dimensionless electric and magnetic tidal Love numbers as follows~\cite{Cardoso:2017cfl}:
\begin{equation}\label{equkk}
\begin{split}
	k_\ell^E &= -\frac{1}{2} \frac{\ell(\ell-1)}{M^{2\ell+1}} \sqrt{\frac{4\pi}{2\ell+1}} \frac{M_\ell}{\mathcal{E}_\ell}\,,\\
    k_\ell^B &= -\frac{3}{2} \frac{\ell(\ell-1)}{(\ell+1) M^{2\ell+1}} \sqrt{\frac{4\pi}{2\ell+1}} \frac{S_\ell}{\mathcal{B}_\ell}\,,
\end{split}
\end{equation}
where $M$ is the mass of MSBSs. $M_\ell$ and $S_\ell$ represent the induced mass and current multipole moments, respectively, while $\mathcal{E}_\ell$ and $\mathcal{B}_\ell$ denote the corresponding electric and magnetic multipole moments of the external tidal field. These multipole moments can be extracted from the asymptotic expansion of the perturbed metric. We adopt the method proposed by K. S. Thorne for defining multipole coefficients of an arbitrary spacetime metric~\cite{Thorne:1980ru}. In the asymptotically Cartesian mass-centered (ACMC) coordinate system, the asymptotic expansions of the $(t,t)$ and $(t,\varphi)$ components of the metric are given by
\begin{equation}\label{equgg}
\begin{split}
	g_{tt} &= -1 + \frac{2M}{r} + \sum_{\ell \geq 2} \left( \frac{2}{r^{\ell+1}} \left[\sqrt{\frac{4\pi}{2\ell+1}} M_\ell Y^{\ell0} + (\ell' < \ell \text{ pole}) \right] - \frac{2}{\ell(\ell-1)}r^\ell \left[ \mathcal{E}_\ell Y^{\ell0} + (\ell' < \ell \text{ pole}) \right] \right)\,,\\
    g_{t\varphi} &= \frac{2J}{r} \sin^2\theta + \sum_{\ell \geq 2} \left( \frac{2}{r^\ell} \left[\sqrt{\frac{4\pi}{2\ell+1}} \frac{S_\ell}{\ell} F^{\ell0}_{\varphi} + (\ell' < \ell \text{ pole}) \right] + \frac{2r^{\ell+1}}{3\ell(\ell-1)} \left[\mathcal{B}_\ell F^{\ell0}_{\varphi} + (\ell' < \ell \text{ pole}) \right] \right)\,.
\end{split}
\end{equation}
In this paper, we focus on $\ell=2$ and calculate the quadrupolar tidal Love numbers.

\subsection{Electric perturbation}

Substituting the perturbed scalar fields given by Eq. (\ref{equfp}) and the even-parity metric perturbation (\ref{equhe}) into the perturbed Einstein field equations (\ref{equEp}), and combining the $(\theta,\theta)$ and $(r,r)$ components of the field equations, we obtain
$H_2(r)=H_0(r)$. From the $(r,\theta)$ component, we have
\begin{equation}\label{equK}
	K'(r) = H_0'(r) + H_0(r) u'(r) - \frac{32\pi \psi_{0}(r) \phi_0'(r)}{r}-\frac{32\pi \psi_1(r) \phi_1'(r)}{r}\,,
\end{equation}
where $\phi_0(r)$ and $\phi_1(r)$ depend on the background solution. By subtracting the $(t,t)$ component of Eq. (\ref{equEp}) from the $(r,r)$ component, and substituting the resulting expressions for $K'(r)$ and $H_2(r)=H_0(r)$ into the resulting equation, we get the equation for $H_0(r)$
\begin{equation}\label{equH}
	a_1 H_0 + a_2 H_0' + H_0'' = a_3 \psi_0+a_4 \psi_1\,,
\end{equation}
where $a_1$, $a_2$, $a_3$ and $a_4$ are given by
\begin{equation}\label{equa}
\begin{split}
	a_1 &= 32\pi e^{v-u} \omega_0^2 \phi_0^2 + 32\pi e^{v-u} \omega_1^2 \phi_1^2 - \frac{6e^v}{r^2} + \frac{2u'}{r} - \frac{u'v'}{2} - \frac{u'^2}{2} + u''\,,\\
    a_2 &= \frac{2}{r} + \frac{u'}{2} - \frac{v'}{2}\,,\\
    a_3 &= \frac{16\pi}{r^2} \left[ -2r e^{v-u} \omega_0^2 \phi_0 + (4 - r u' - r v') \phi_0' + 2r \phi_0'' \right]\,,\\
    a_4 &= \frac{16\pi}{r^2} \left[ -2r e^{v-u} \omega_1^2 \phi_1 + (4 - r u' - r v') \phi_1' + 2r \phi_1'' \right]\,.
\end{split}
\end{equation}
Similarly, from the perturbed matter field equation (\ref{equMp}), we obtain the equation for $\psi_0$
\begin{equation}\label{equpsi0}
	b_1 \psi_0 + b_2 \psi_0' + \psi_0'' = b_3 H_0+b_4\psi_1\,,
\end{equation}
where $b_1$, $b_2$, $b_3$ and $b_4$ are given by
\begin{equation}\label{equb}
\begin{split}
	b_1 &= -\frac{6 e^{v}}{r^2} - e^{v} \mu_0^2 + e^{v-u} \omega_0^2 + \frac{v'-u'}{2r} - 32\pi \phi_0'^2\,,\\
    b_2 &= \frac{u'}{2} - \frac{v'}{2}\,,\\
    b_3 &= r[-e^{v-u} \omega_0^2 \phi_0 + \frac{2 \phi_0'}{r} - \frac{(u'+v')\phi_0'}{2} + \phi_0'']\,,\\
    b_4 &= 32\pi\phi_0'\phi_1'
\end{split}
\end{equation}
The equation for $\psi_1$ reads
\begin{equation}\label{equpsi1}
	c_1 \psi_1 + c_2 \psi_1' + \psi_1'' = c_3 H_0+c_4\psi_0\,,
\end{equation}
where $c_1$, $c_2$, $c_3$ and $c_4$ are given by
\begin{equation}\label{equc}
\begin{split}
	c_1 &= -\frac{6 e^{v}}{r^2} - e^{v} \mu_1^2 + e^{v-u} \omega_1^2 + \frac{v'-u'}{2r} - 32\pi \phi_1'^2\,,\\
    c_2 &= \frac{u'}{2} - \frac{v'}{2}\,,\\
    c_3 &= r[-e^{v-u} \omega_1^2 \phi_1 + \frac{2 \phi_1'}{r} - \frac{(u'+v')\phi_1'}{2} + \phi_1'']\,,\\
    c_4 &= 32\pi\phi_0'\phi_1'
\end{split}
\end{equation}

When calculating the Love numbers, we perform the calculations far away from the center of the MSBSs. Therefore, Eq. (\ref{equH}) can be simplified to
\begin{equation}\label{equHs}
	H_0'' + \left( \frac{2}{r} + \frac{u' - v'}{2} \right) H_0' + \left( -\frac{6e^{v}}{r^2} + \frac{2u'}{r} - \frac{u'^2}{2} - \frac{u'v'}{2} + u'' \right) H_0 = 0\,.
\end{equation}
Substituting $e^{-v(r)}=1-2M/r$ into Eq. (\ref{equHs}) yields
\begin{equation}\label{equHf}
	H_0'' + \frac{2(r - M)}{r(r - 2M)} H_0' + \frac{6r(2M - r) - 4M^2}{r^2 (r - 2M)^2} H_0 = 0\,.
\end{equation}
This equation has a general solution
\begin{equation}\label{equHsol}
	H_0 = E_p P_2^2\left(\frac{r}{M} - 1\right) + E_q Q_2^2\left(\frac{r}{M} - 1\right)\,,
\end{equation}
where $P_2^2$ and $Q_2^2$ are the associated Legendre functions of the first and second kind, respectively. Using the asymptotic behavior of $H_0$ and comparing with Eq. (\ref{equgg}), the electric tidal Love numbers can be obtained.
In the actual calculation of the Love numbers, we define
\begin{equation}\label{equyH}
	y = R_{\text{ext}} \frac{H_0'(R_{\text{ext}})}{H_0(R_{\text{ext}})}\,,
\end{equation}
where $R_{\text{ext}}$ is the extraction radius. Then, the electric tidal Love numbers are obtained, which read
\begin{equation}\label{equkE}
\begin{split}
k_2^E &= \frac{8}{5} (1 - 2C)^2 \bigl[ 2C(y - 1) - y + 2 \bigr] \\
&\times \Bigg( 2C \Big[ 4(y + 1)C^4 + (6y - 4)C^3 + (26 - 22y)C^2  + 3(5y - 8)C - 3y + 6 \Big] \\
&\qquad - 3(1 - 2C)^2 \bigl( 2C(y - 1) - y + 2 \bigr) \log \frac{1}{1 - 2C} \Bigg)^{-1}\,,
\end{split}
\end{equation}
where $C=M/R_{\text{ext}}$.

\subsection{Magnetic perturbation}

Substituting Eqs. (\ref{equho}) and (\ref{equfp}) into the perturbed Einstein field equations (\ref{equEp}), and using the $(r, \theta)$ component of Eq. (\ref{equEp}), we obtain $\psi_0=\psi_1=0$. Meanwhile, the $(t,\varphi)$ component of the perturbed Einstein equations yields the equation for $h_0$
\begin{equation}\label{equh0}
	d_1 h_0 + d_2 h_0' + h_0'' = 0\,,
\end{equation}
where $d_1$ and $d_2$ are given by
\begin{equation}\label{equd}
\begin{split}
	d_1 &= \frac{-4e^{v} + r (u' + v') - 2}{r^2}\,,\\
    d_2 &= -\frac{u' + v'}{2}\,.
\end{split}
\end{equation}

We extract the tidal Love numbers far away from the center of the MSBSs. Therefore, Eq. (\ref{equh0}) can be simplified to
\begin{equation}\label{equhf}
	 \frac{4M - 6r}{r^2(r - 2M)} h_0'(r) + h_0''(r) = 0\,.
\end{equation}
The solution to this equation can be expressed in terms of elementary functions. By comparing the asymptotic behavior of the solution with Eq. (\ref{equgg}), the magnetic tidal Love numbers can be obtained.
Similarly, we define
\begin{equation}\label{equyh}
	y = R_{\text{ext}} \frac{h_0'(R_{\text{ext}})}{h_0(R_{\text{ext}})}\,,
\end{equation}
Then, the magnetic tidal Love numbers are calculated to be
\begin{equation}\label{equkB}
	k_2^B = \frac{8}{5} \frac{2C(y - 2) - y + 3}{2C \Big[ 2C^3(y + 1) + 2C^2 y + 3C(y - 1) - 3y + 9 \Big] + 3 \bigl[ 2C(y - 2) - y + 3 \bigr] \log(1 - 2C)}\,,
\end{equation}
where $C=M/R_{\text{ext}}$.

\section{BACKGROUND SOLUTIONS}\label{Sec4}

In this section, we present the numerical results for the background solutions of MSBSs. In order to solve the ordinary differential Eqs. (\ref{equu})-(\ref{equf1}), suitable boundary conditions must be set. Given that the spacetime is regular and asymptotically flat, and the scalar fields vanish at infinity, the boundary conditions for the metric and field functions are given by
\begin{equation}\label{equbakb}
\begin{split}
u(0) = u_c\,,& \qquad v(0) = 0\,, 
\\ \phi_n(0) = \phi_{nc}\,,& \qquad \phi_n'(0) = 0\,,\quad n =0,1\,,
\end{split}
\end{equation}
where $u_c$ is arbitrary. At infinity, the boundary conditions are
\begin{equation}\label{equbakbinf}
   \lim_{r \to \infty} u(r) = 0\,, \qquad \lim_{r \to \infty} \phi_n(r) = 0\,,\quad n =0,1\, . 
\end{equation}
The ADM mass of MSBSs is $M=m(r\to \infty)$, which can be obtained from the following equation
\begin{equation}\label{equmr}
   m(r) = \frac{r}{2} \left(1 - e^{-v(r)} \right)\, .
\end{equation}
Since the scalar fields are distributed throughout space and decay exponentially, MSBSs have no rigid surface. Therefore, we take the radius enclosing 99\% of the total mass of MSBSs as the effective radius $R$. 

In numerical calculations, we adopt the following dimensionless transformations
\begin{equation}\label{equtrab}
\begin{split}
\tilde{r} = r/\rho, \quad \tilde{\omega}_n = \omega_n\rho, \quad \tilde{\mu}_n = \mu_n\rho, \quad n=0,1\,,
\end{split}
\end{equation}
where $\rho$ is a positive constant with dimension of length, and we set $\rho=1/\mu_0$. Furthermore, we define a new radial coordinate $x$ as
\begin{equation}\label{equtrar}
x = \frac{\tilde{r}}{1+\tilde{r}}\,, 
\end{equation}
where the radial coordinate $\tilde{r}\in[0, \infty)$, so $x\in[0, 1]$.
Using the finite element method, we solve the differential equations on a grid of 10000 points in the interval $0 \le x \le 1$. To ensure the accuracy of the results, we require the relative error to be less than $10^{-5}$.

For convenience, we denote the boson stars in the ground state and the first excited state as $B_0$ and $B_1$, respectively, and the MSBSs as $B_0B_1$. We solve for the background solutions of MSBSs under both synchronized and nonsynchronized frequency conditions. Synchronized frequency corresponds to $\omega_0 = \omega_1$ with $\mu_0 \ne \mu_1$. Nonsynchronized frequency corresponds to $\omega_0 \ne \omega_1$ with $\mu_0 = \mu_1$. We now present the different solutions in detail.

\subsection{Synchronized frequency}

Under synchronized frequency conditions, we vary the mass ratio $\tilde{\mu}_1=\mu_1/\mu_0$. The solutions can be classified into two types: single-branch and double-branch. In the single-branch case, each frequency corresponds to one solution. In the double-branch case, when the frequency takes values in some ranges, each frequency corresponds to two distinct solutions. Based on the numerical results, the solutions are single-branch for $0.7976 \le \tilde{\mu}_1 < 1$, and double-branch for $0.7922 \le \tilde{\mu}_1 < 0.7976$.

\subsubsection{Single-branch}

\begin{figure}[!htbp]
		\begin{center}
		\subfigure{ 
			\includegraphics[height=.28\textheight,width=.32\textheight, angle =0]{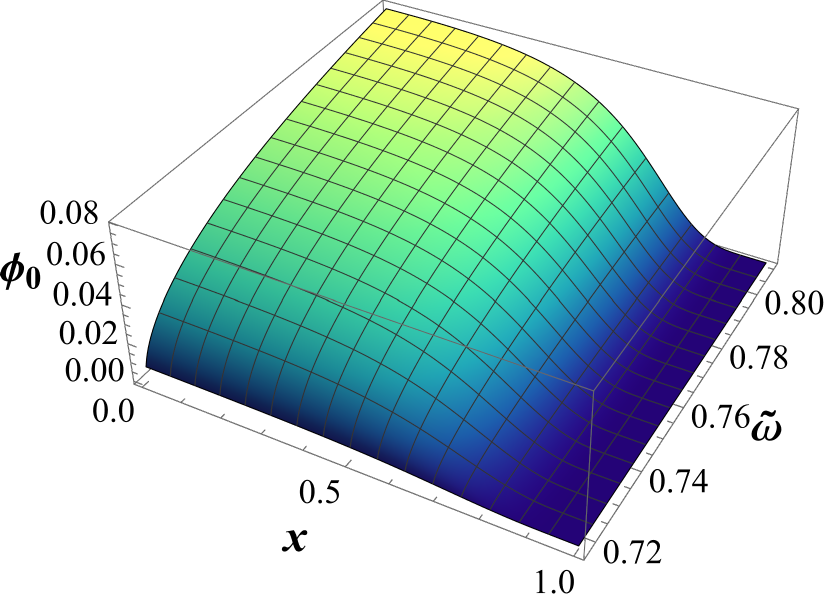}
			\label{fig:s0.853f0}
		}	 
  		\subfigure{  
			\includegraphics[height=.28\textheight,width=.32\textheight, angle =0]{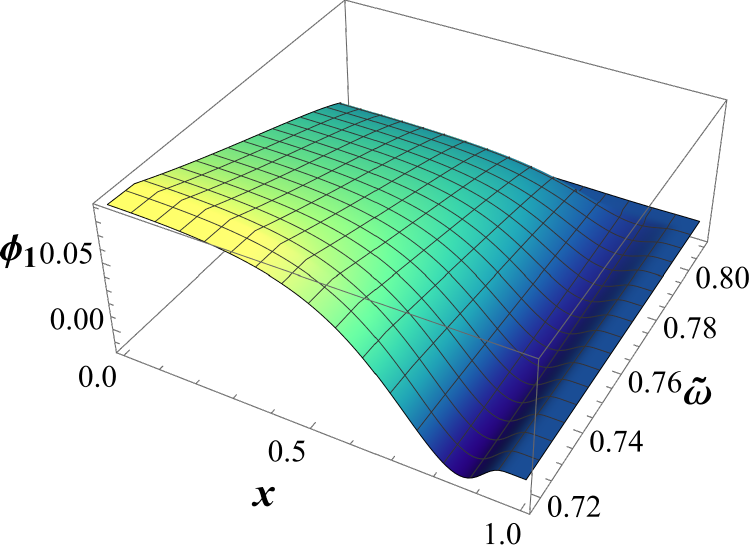}
			\label{fig:s0.853f1}
		}	    
  		\end{center}		
		\caption{ The matter field functions $\phi_0$ and $\phi_1$ as functions of $x$ and $\tilde{\omega}$ for $\tilde{\mu}_1=0.853$ and $\tilde{\mu}_0=1$. }
	\label{fig:sf}	
		\end{figure}

We first discuss the case of synchronized frequency single-branch solutions. Fig.~\ref{fig:sf} shows the matter field functions $\phi_0$ and $\phi_1$ at
different synchronized frequencies, for $\tilde{\mu}_1=0.853$ and $\tilde{\mu}_0=1$. It can be seen from the figure that the ground state scalar field function $\phi_0$ has no radial node, while the first excited state scalar field function $\phi_1$ has one node. As the frequency $\tilde{\omega}$ increases, $\phi_0$ increases, whereas $\phi_1$ decreases. At the minimum frequency, the ground state field vanishes and the excited state field reaches its maximum. Conversely, at the maximum frequency, the ground state field reaches its maximum, while the first excited state field tends to disappear.

\begin{figure}[!htbp]
		\begin{center}
		\subfigure{ 
        \includegraphics[height=.24\textheight, angle =0]{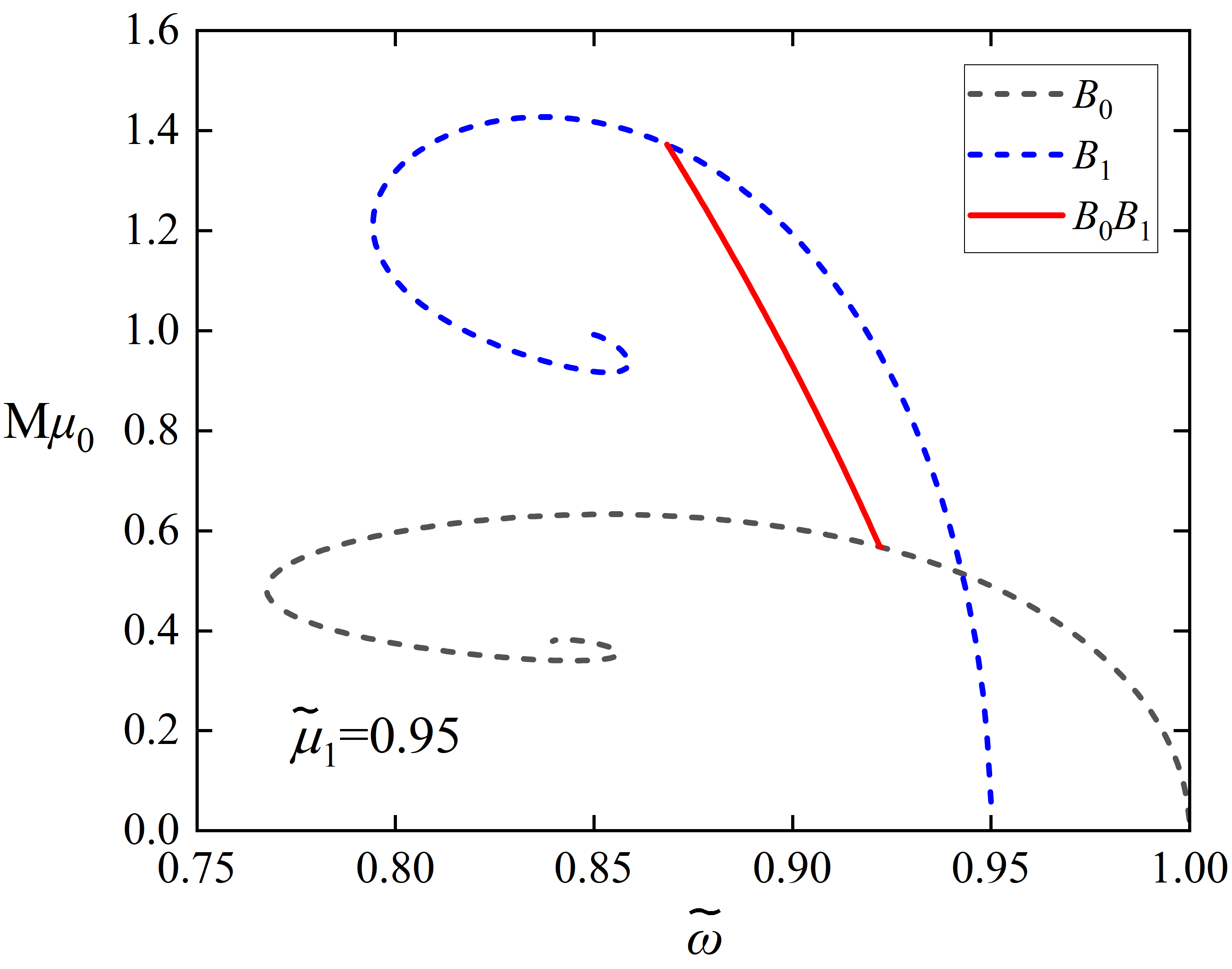}
			\label{fig:ms0.95}
		}	 
  		\subfigure{  
			\includegraphics[height=.24\textheight, angle =0]{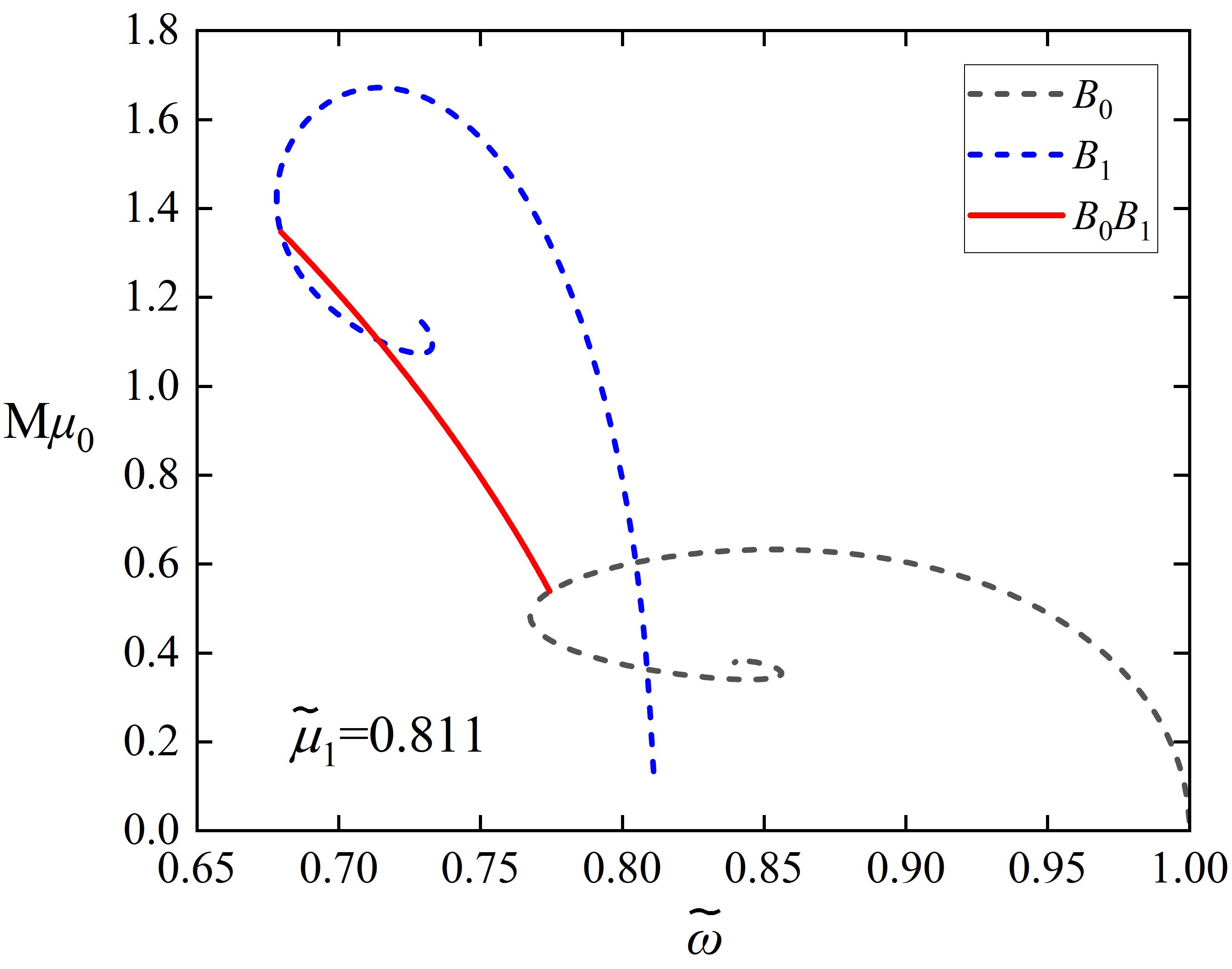}
			\label{fig:ms0.811}
		}	
        \subfigure{  
			\includegraphics[height=.24\textheight, angle =0]{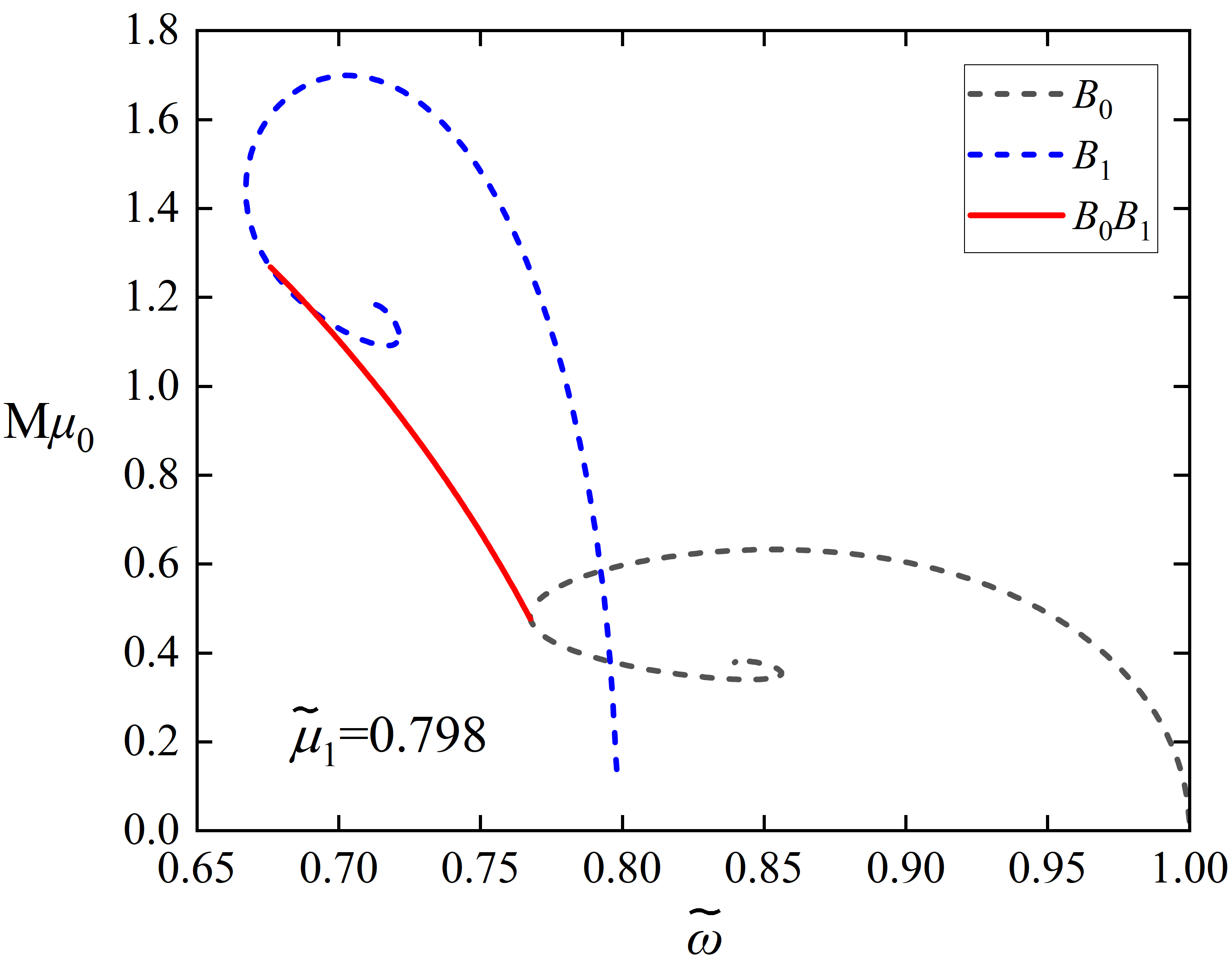}
			\label{fig:ms0.798}
		}
        \subfigure{  
			\includegraphics[height=.24\textheight, angle =0]{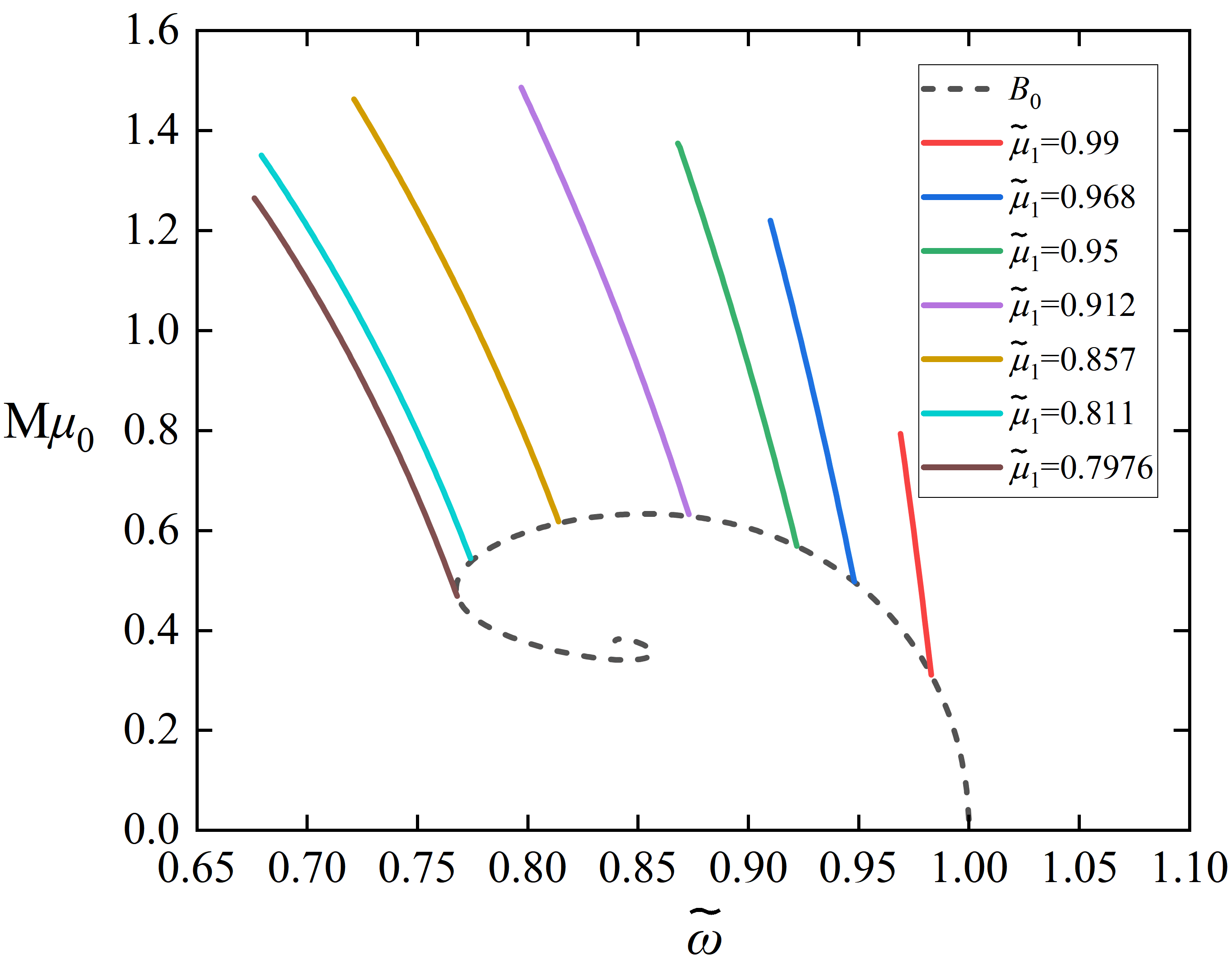}
			\label{fig:msall}
		}        
  		\end{center}		
		\caption{ The ADM mass $M$ of the MSBSs as a function of the synchronized frequency $\tilde{\omega}$ for  several values of $\tilde{\mu}_1$. All solutions correspond to $\tilde{\mu}_0=1$. }
	\label{fig:ms}	
		\end{figure}

Then, we analyze the relationship between the ADM mass $M$ and the synchronized frequency $\tilde{\omega}$. In Fig.~\ref{fig:ms}, the black dashed lines $(B_0)$ represent the solutions of the ground state boson stars with $\tilde{\mu}_0=1$. The blue dashed lines $(B_1)$ in the first three panels represent the first excited state boson stars with $\tilde{\mu}_1=0.95, 0.811, 0.798$, respectively. All solid lines correspond to the MSBSs for different values of $\tilde{\mu}_1$ with $\tilde{\mu}_0=1$. It can be clearly seen from the figure that the $M-\omega$ curves of the MSBSs lie in the region between the $M-\omega$ curves of the ground state and the first excited state boson stars. The figure also shows that the mass $M$ decreases as $\tilde{\omega}$ increases. The red solid lines intersect the blue dashed lines at one end and the black dashed lines at the other end. This means that the MSBSs degenerate into the first excited state boson stars at the minimum frequency and into the ground state boson stars at the maximum frequency. This behavior is also reflected in Fig.~\ref{fig:sf}, where the ground state and first excited state field functions of the MSBSs vanish accordingly. From the fourth panel, we can see that as $\tilde{\mu}_1$ decreases, the $M-\omega$ curves of the MSBSs shift from right to left. Observing the intersections with the black dashed line, we find that when $\tilde{\mu}_1=0.7976$, the MSBSs can degenerate into the second branch of the ground state boson stars.

%%%%%%%%%%%%%%%%%%%%%%%%%%%%%%%%%%%%%%%%%%%%%%%%%%%%%%%%%%%
%%%%%%%%%%%%%%%%%%%%%%%%%%%%%%%%%%%%%%%%%%%%%%%%%%%%%%%%%%%        

%%%%%%%%%%%%%%%%%%%%%%%%%%%%%%%%%%%%%%%%%%%%%%%%%%%%%%%%%%%		
  %%%%%%%%%%%%%%%%%%%%%%%%%%%%%%%%%%%%%%%%%%%%%%%%%%%%%%%%%%%
\subsubsection{Double-branch}

\begin{figure}[!htbp]
		\begin{center}
		\subfigure{ 
        \includegraphics[height=.28\textheight,width=.32\textheight, angle =0]{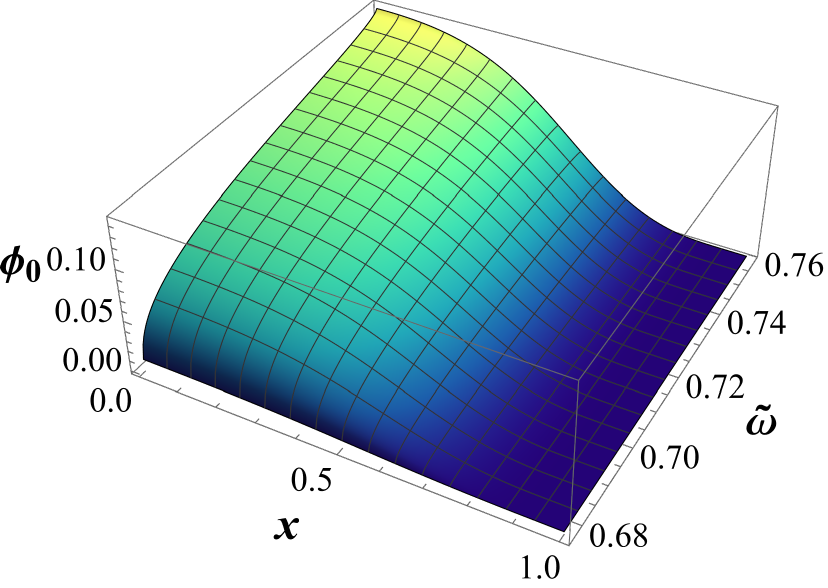}
			\label{fig:d0.797f01}
		}	 
  		\subfigure{  
			\includegraphics[height=.28\textheight,width=.32\textheight, angle =0]{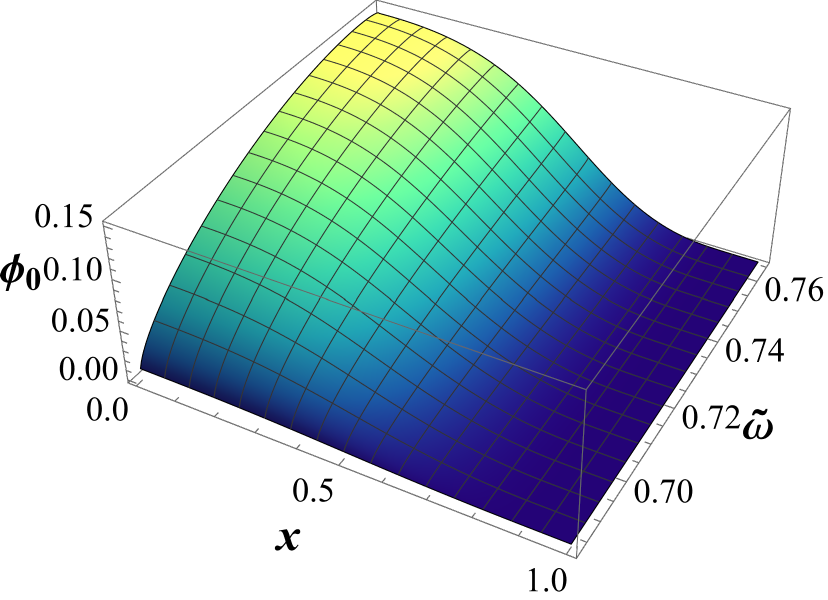}
			\label{fig:d0.797f02}
		}	
        \subfigure{  
			\includegraphics[height=.28\textheight,width=.32\textheight, angle =0]{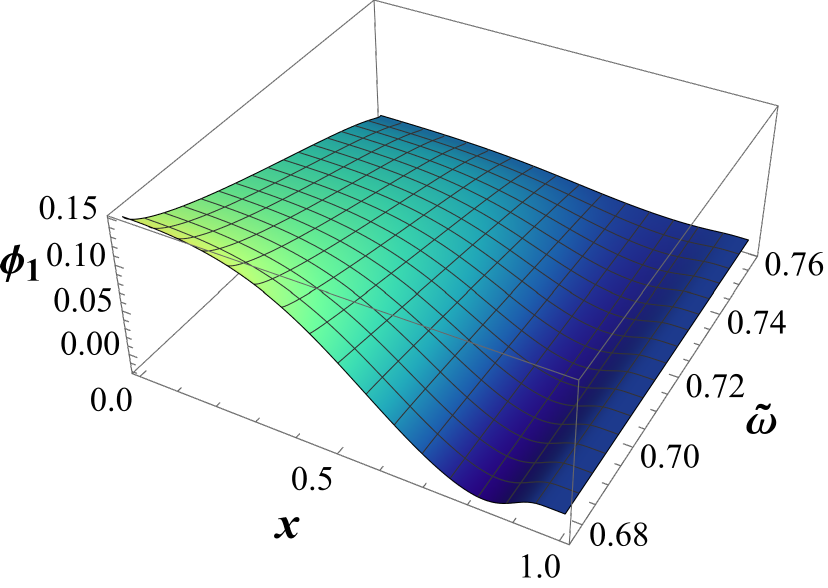}
			\label{fig:d0.797f11}
		}
        \subfigure{  
			\includegraphics[height=.28\textheight,width=.32\textheight, angle =0]{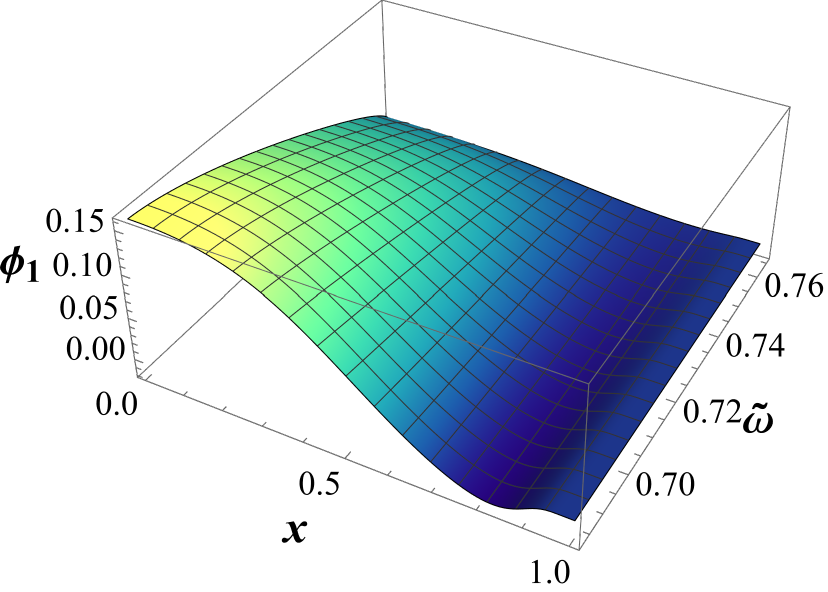}
			\label{fig:d0.797f12}
		}        
  		\end{center}		
		\caption{ The matter field functions $\phi_0$ and $\phi_1$ on the first (left two panels) and second (right two panels) branches of the MSBSs as functions of $x$ and $\tilde{\omega}$ for $\tilde{\mu}_1=0.797$ and $\tilde{\mu}_0=1$. }
	\label{fig:df7}	
		\end{figure}

\begin{figure}[!htbp]
		\begin{center}
		\subfigure{ 
        \includegraphics[height=.28\textheight,width=.32\textheight, angle =0]{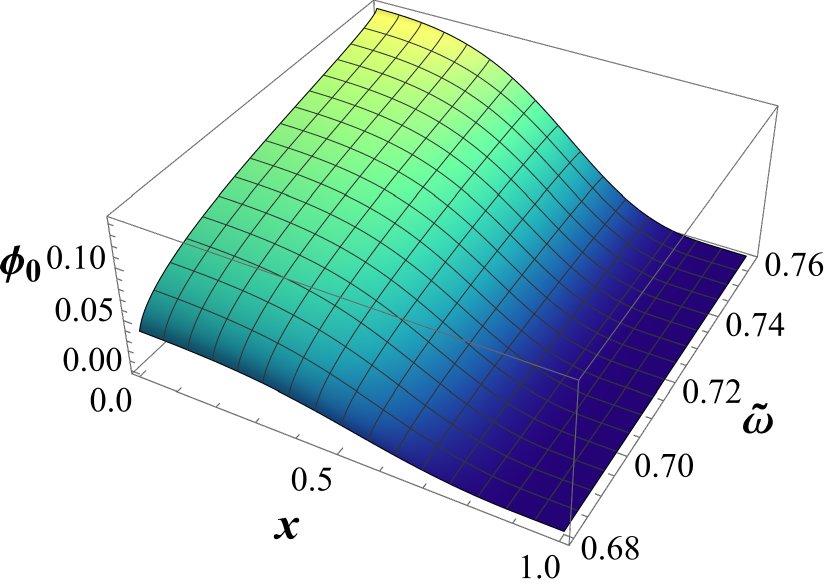}
			\label{fig:d0.796f01}
		}	 
  		\subfigure{  
			\includegraphics[height=.28\textheight,width=.32\textheight, angle =0]{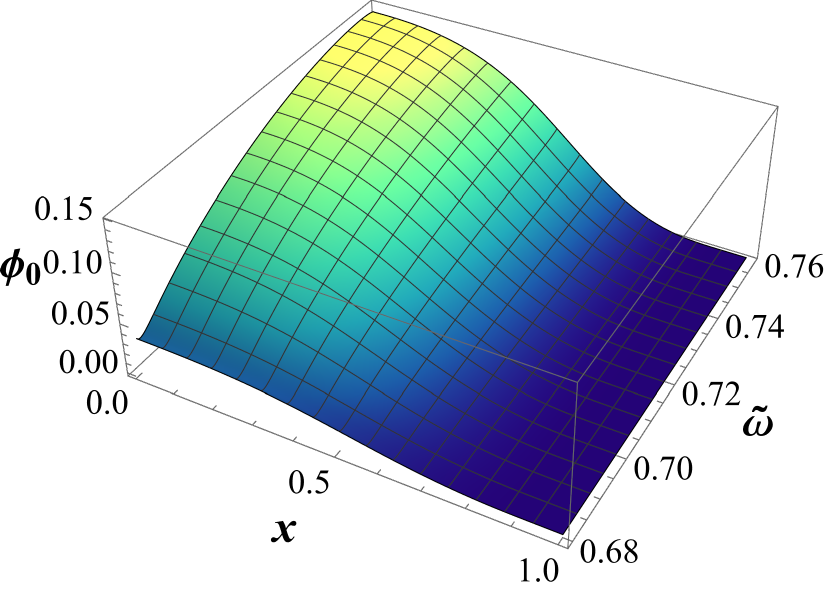}
			\label{fig:d0.796f02}
		}	
        \subfigure{  
			\includegraphics[height=.28\textheight,width=.32\textheight, angle =0]{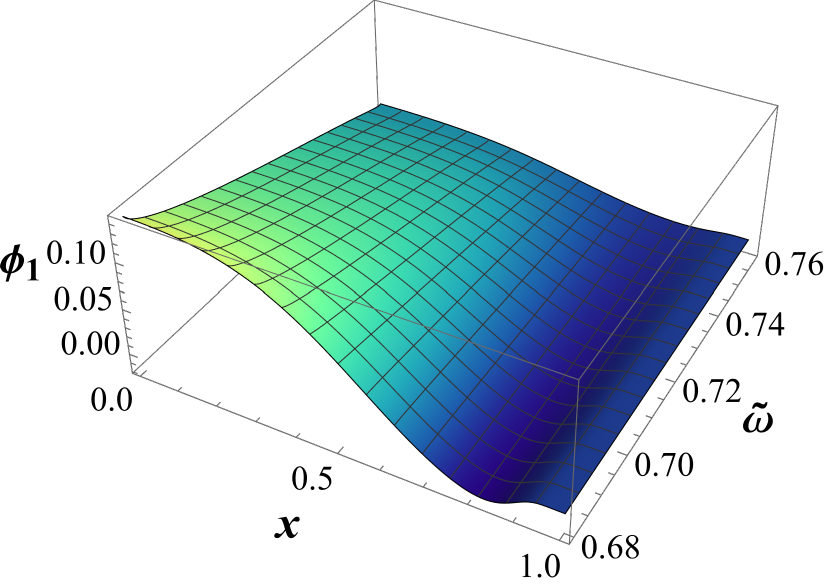}
			\label{fig:d0.796f11}
		}
        \subfigure{  
			\includegraphics[height=.28\textheight,width=.32\textheight, angle =0]{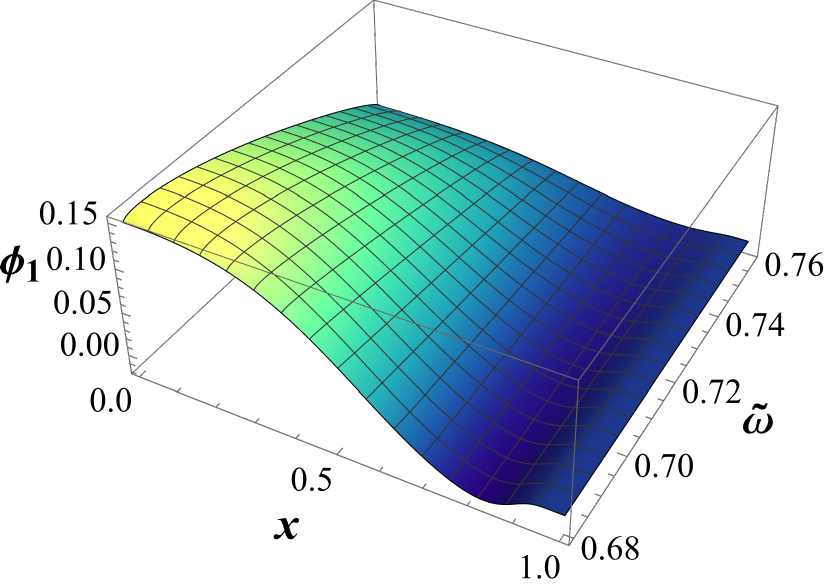}
			\label{fig:d0.796f12}
		}        
  		\end{center}		
		\caption{ The matter field functions $\phi_0$ and $\phi_1$ on the first (left two panels) and second (right two panels) branches of the MSBSs as functions of $x$ and $\tilde{\omega}$ for $\tilde{\mu}_1=0.796$ and $\tilde{\mu}_0=1$. }
	\label{fig:df6}	
		\end{figure}

In the synchronized frequency case, the solutions exhibit two branches for $0.7922 \le \tilde{\mu}_1 < 0.7976$. In Fig.~\ref{fig:df7}, the matter field functions $\phi_0$ and $\phi_1$ are plotted with respect to $x$ and $\tilde{\omega}$ for $\tilde{\mu}_1=0.797$ and $\tilde{\mu}_0=1$. The two left panels represent the first branch of the solutions. As the synchronized frequency increases, $\phi_0$ gradually increases while $\phi_1$ decreases. The second branch of the solutions is shown in the two right panels. The trends of $\phi_0$ and $\phi_1$ with frequency are similar to the single-branch case, but the peak values differ. It can be seen from the upper two panels that the ground state field gradually vanishes as the frequency decreases. However, from the lower two panels, it is observed that the excited state field always exists as the frequency increases. Fig.~\ref{fig:df6} shows the matter field functions for $\tilde{\mu}_1=0.796$ and $\tilde{\mu}_0=1$. The variation trends are almost the same as those for $\tilde{\mu}_1=0.797$, but the difference is that when the synchronized frequency approaches the minimum, the ground state fields of both the first and second branches do not vanish.

\begin{figure}[!htbp]
		\begin{center}
		\subfigure{ 
        \includegraphics[height=.24\textheight, angle =0]{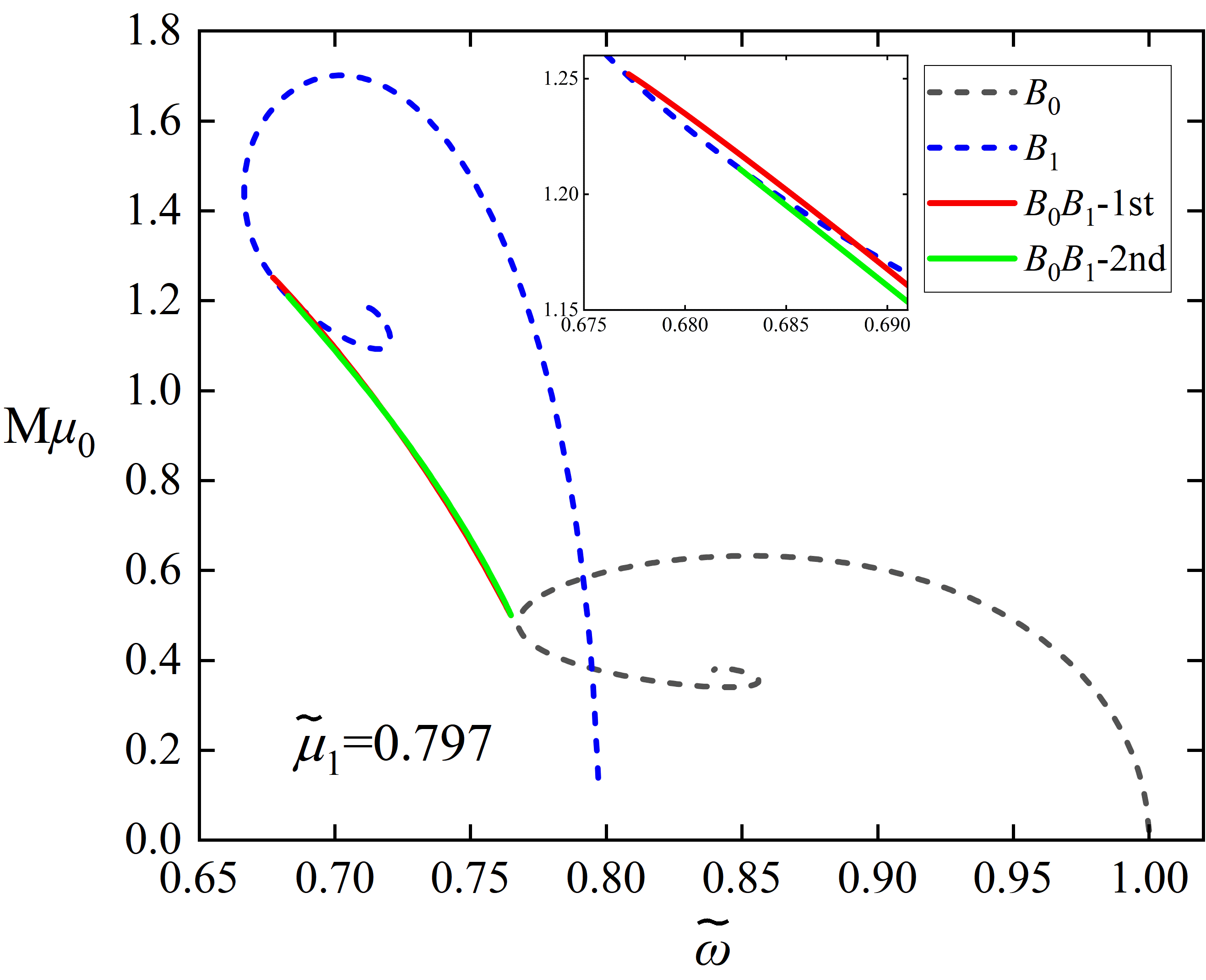}
			\label{fig:md0.797}
		}	 
  		\subfigure{  
			\includegraphics[height=.24\textheight, angle =0]{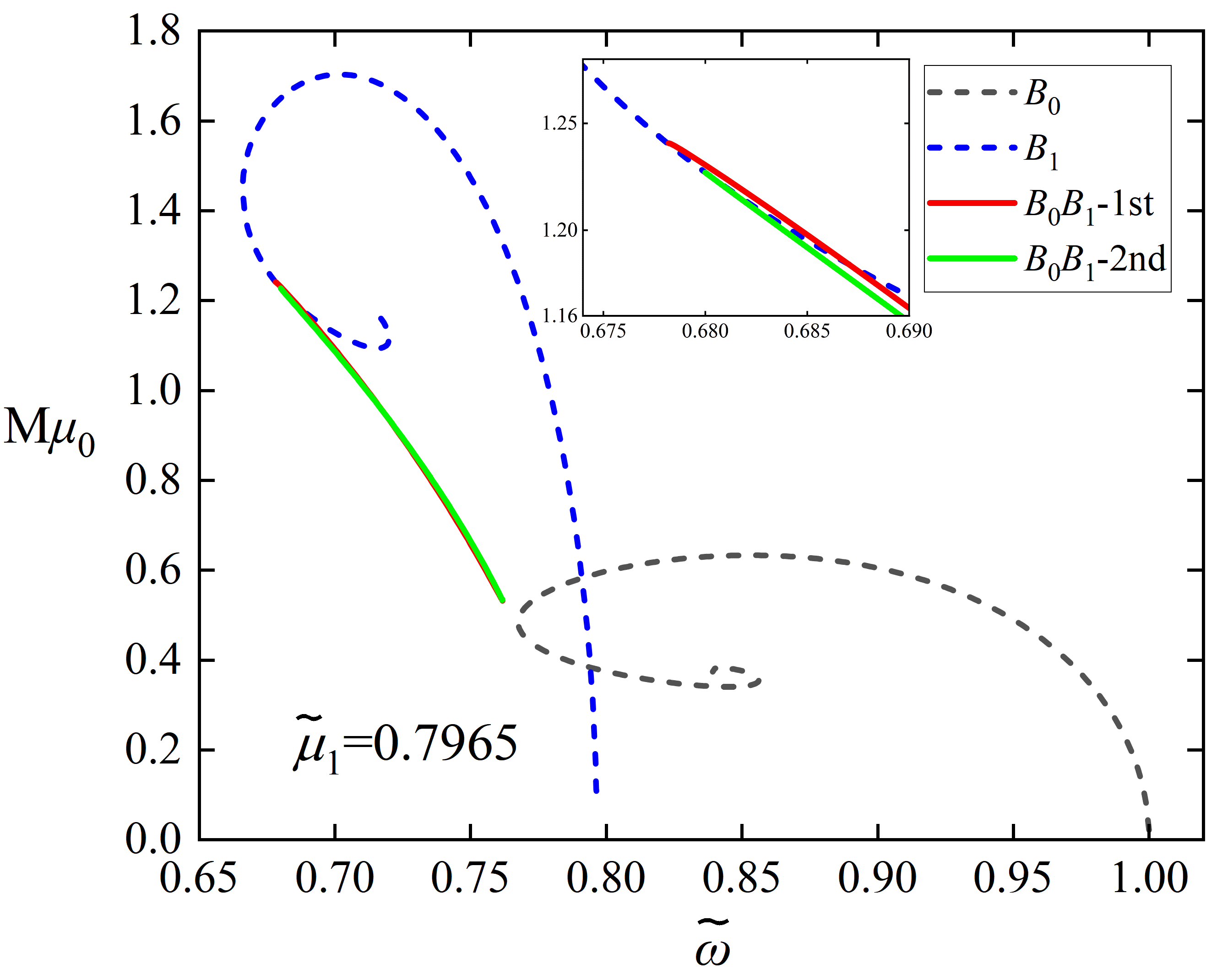}
			\label{fig:md0.7965}
		}	
        \subfigure{  
			\includegraphics[height=.24\textheight, angle =0]{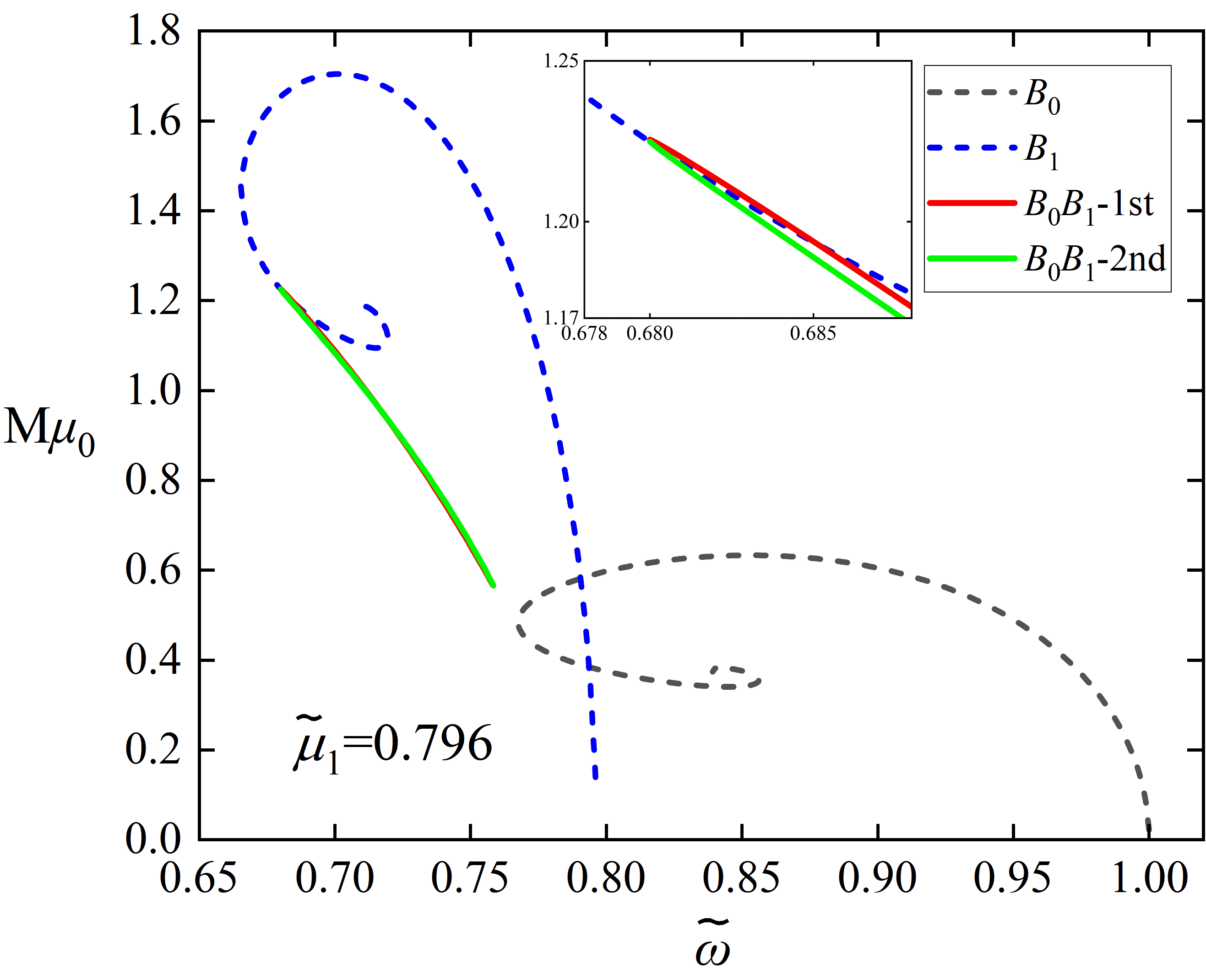}
			\label{fig:md0.796}
		}
        \subfigure{  
			\includegraphics[height=.24\textheight, angle =0]{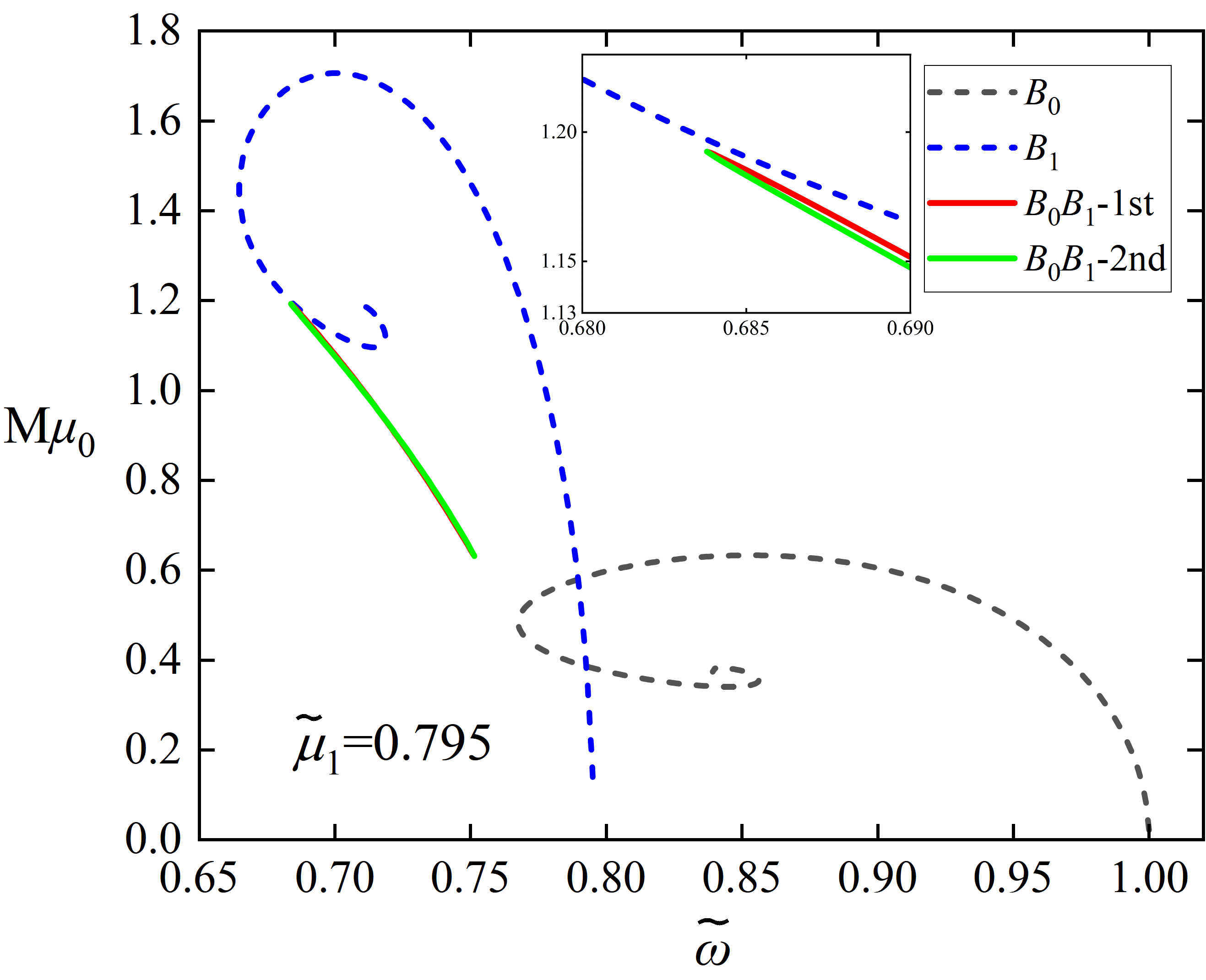}
			\label{fig:md0.795}
		} 
        \subfigure{  
			\includegraphics[height=.24\textheight, angle =0]{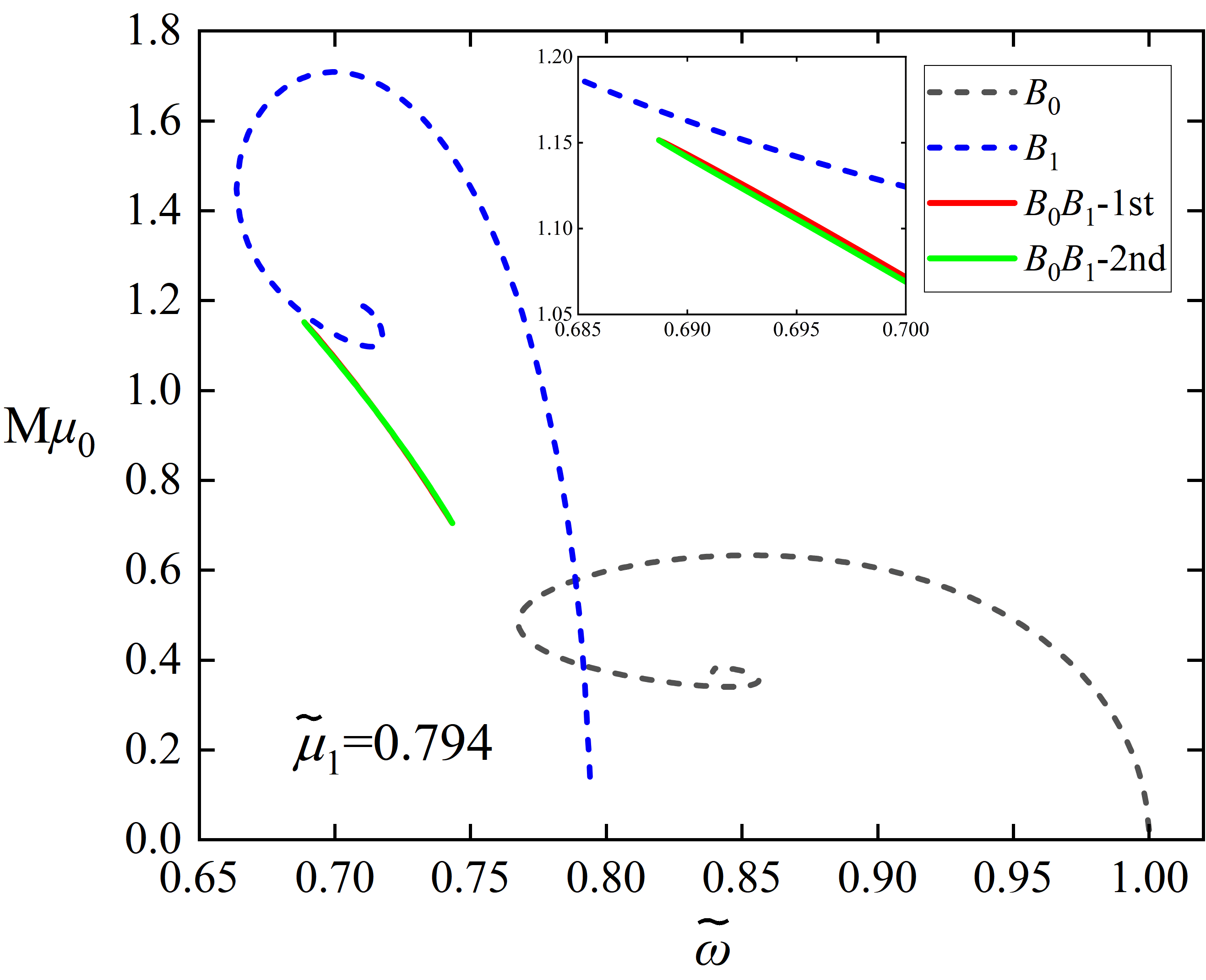}
			\label{fig:md0.794}
		}
        \subfigure{  
			\includegraphics[height=.24\textheight, angle =0]{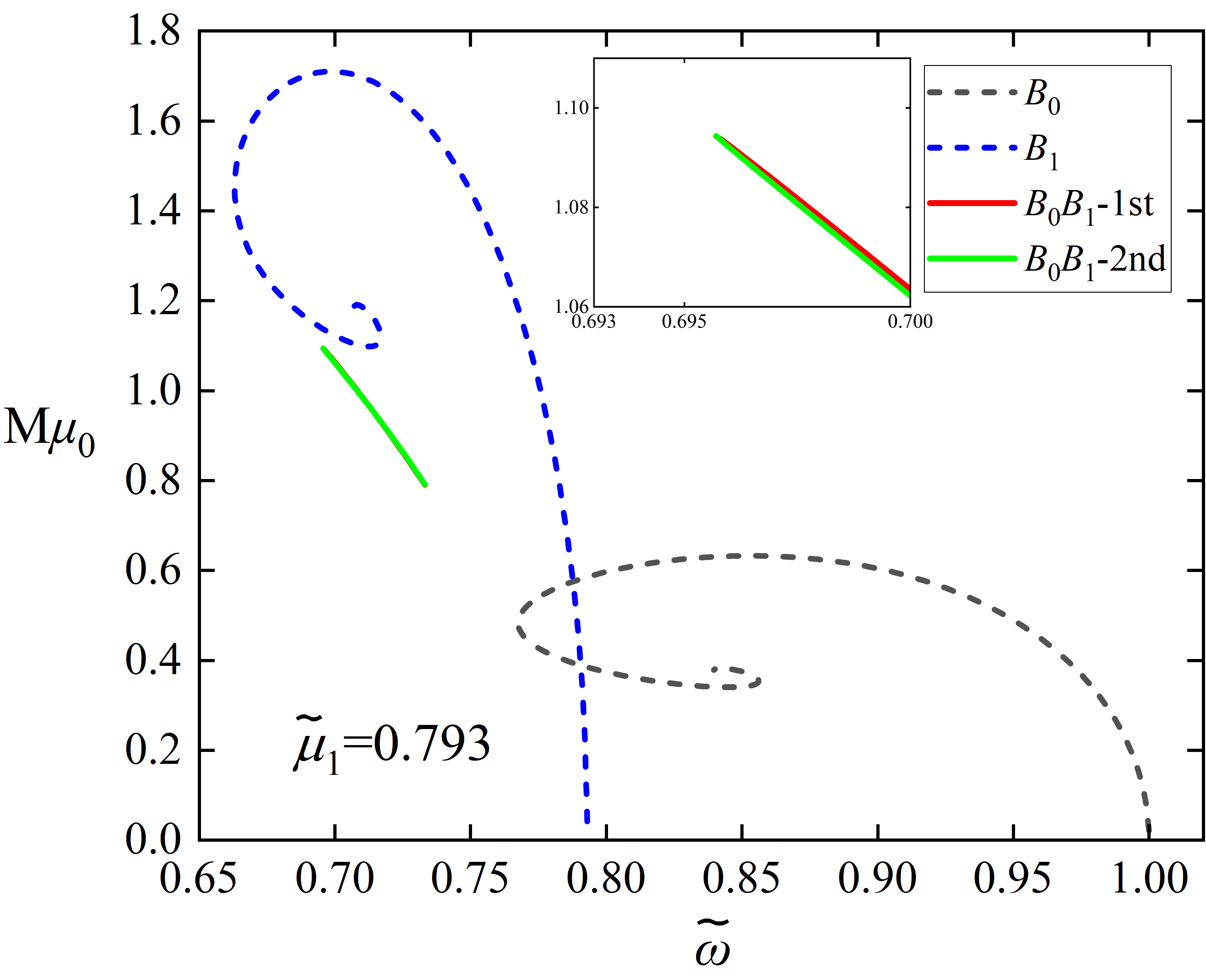}
			\label{fig:md0.793}
		}
  		\end{center}		
		\caption{ The ADM mass $M$ of the MSBSs as a function of the synchronized frequency $\tilde{\omega}$ for $\tilde{\mu}_1=0.797, 0.7965, 0.796, 0.795, 0.794, 0.793$ and $\tilde{\mu}_0=1$. }
	\label{fig:md}	
		\end{figure}

Fig.~\ref{fig:md} shows the ADM mass $M$ against the synchronized frequency $\tilde{\omega}$. The black and blue dashed lines represent $B_0$ and $B_1$, respectively. The red and green solid lines represent the first and second branches of $B_0B_1$, respectively. In the MSBSs, there appears a bifurcation phenomenon, the same as in multi-state Dirac stars, where the single-branch solutions cannot continuously transition to the double-branch solutions~\cite{Liang:2023ywv}. As shown in the figure, the two branches are very close to each other. From the panel for $\tilde{\mu}_1=0.797$, the intersections of the first and second branches with the blue dashed line are clearly different. As $\tilde{\mu}_1$ decreases, these two intersections gradually approach each other. When $\tilde{\mu}_1$ drops below 
$0.796$, the two branches merge at both ends. In the panels for $\tilde{\mu}_1=0.795,0.794,0.793$, the solutions do not intersect the blue dashed line at the minimum frequency, meaning that the MSBSs cannot degenerate into the first excited state boson stars. In fact, for $\tilde{\mu}_1=0.796$, although $B_0B_1$ intersects $B_1$ at the minimum frequency, Fig.~\ref{fig:df6} shows that the ground state field does not vanish. This indicates that the MSBSs do not actually degenerate into the excited state boson stars. In addition, the existence range of the synchronized frequency for MSBSs decreases as $\tilde{\mu}_1$ decreases.
%%%%%%%%%%%%%%%%%%%%%%%%%%%%%%%%%%%%%%%%%%%%%%%%%%%%%%%%%%%		
  %%%%%%%%%%%%%%%%%%%%%%%%%%%%%%%%%%%%%%%%%%%%%%%%%%%%%%%%%%%

  %%%%%%%%%%%%%%%%%%%%%%%%%%%%%%%%%%%%%%%%%%%%%%%%%%%%%%%%%%%		
  %%%%%%%%%%%%%%%%%%%%%%%%%%%%%%%%%%%%%%%%%%%%%%%%%%%%%%%%%%%

 %%%%%%%%%%%%%%%%%%%%%%%%%%%%%%%%%%%%%%%%%%%%%%%%%%%%%%%%%%%		
  %%%%%%%%%%%%%%%%%%%%%%%%%%%%%%%%%%%%%%%%%%%%%%%%%%%%%%%%%%%
  
\subsection{Nonsynchronized frequency}

In this section, we discuss the MSBSs in the case of nonsynchronized frequency. We set the field masses to $\tilde{\mu}_0=\tilde{\mu}_1=1$. In the numerical calculations, we fix the ground state scalar field frequency $\tilde{\omega}_0$ and investigate the behavior of the solutions versus the excited state scalar field frequency $\tilde{\omega}_1$. Similar to the synchronized frequency case, our numerical results show that the solutions can also be classified into single-branch and double-branch types. Solutions are single-branch for $0.7675 \le \tilde{\omega}_0 < 1$, and double-branch for $0.7214 \le \tilde{\omega}_0 < 0.7675$.

\subsubsection{Single-branch}

\begin{figure}[!htbp]
		\begin{center}
		\subfigure{ 
        \includegraphics[height=.28\textheight,width=.32\textheight, angle =0]{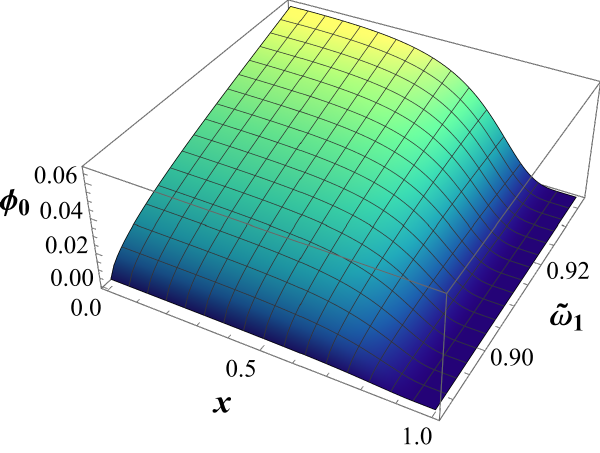}
			\label{fig:ns0.887f0}
		}	 
  		\subfigure{  
			\includegraphics[height=.28\textheight,width=.32\textheight, angle =0]{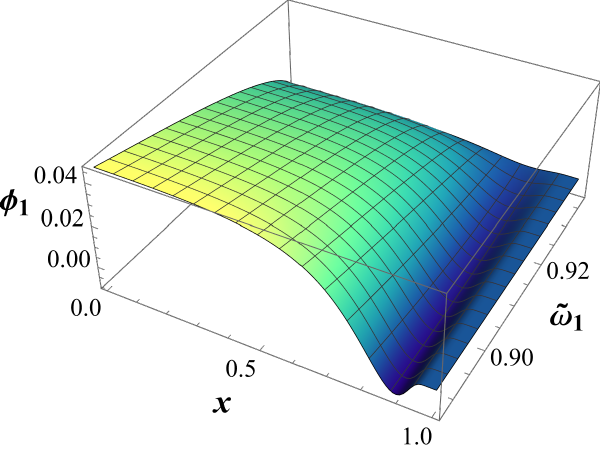}
			\label{fig:ns0.887f1}
		}	        
  		\end{center}		
		\caption{ The matter field functions $\phi_0$ and $\phi_1$ as functions of $x$ and $\tilde{\omega}_1$ for $\tilde{\omega}_0=0.887$ and $\tilde{\mu}_0=\tilde{\mu}_1=1$. }
	\label{fig:nsf}	
		\end{figure}

Fig.~\ref{fig:nsf} shows the matter field functions with the ground state field frequency $\tilde{\omega}_0$ fixed at $0.887$. As shown in the figure, the ground state field function $\phi_0$ has no node in the radial direction, while the first excited state field function $\phi_1$ has one node. As the first excited state field frequency $\tilde{\omega}_1$ increases, $\phi_1$ gradually decreases, and $\phi_0$ gradually increases. Furthermore, as the frequency $\tilde{\omega}_1$ approaches the minimum, $\phi_0$ tends to zero and $\phi_1$ reaches its maximum; the opposite occurs at the maximum frequency.

\begin{figure}[!htbp]
		\begin{center}
		\subfigure{ 
        \includegraphics[height=.24\textheight, angle =0]{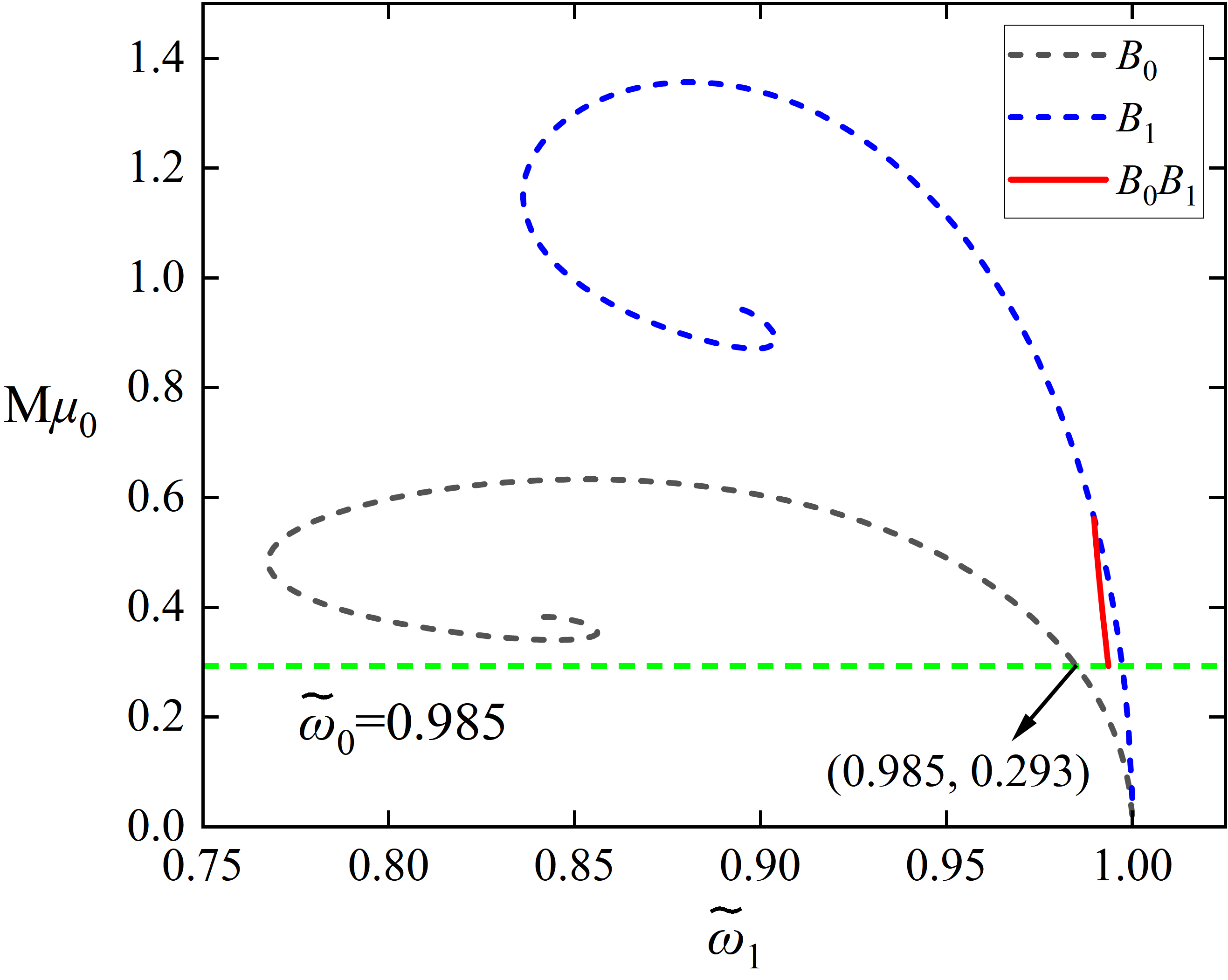}
			\label{fig:nms0.985}
		}	 
  		\subfigure{  
			\includegraphics[height=.24\textheight, angle =0]{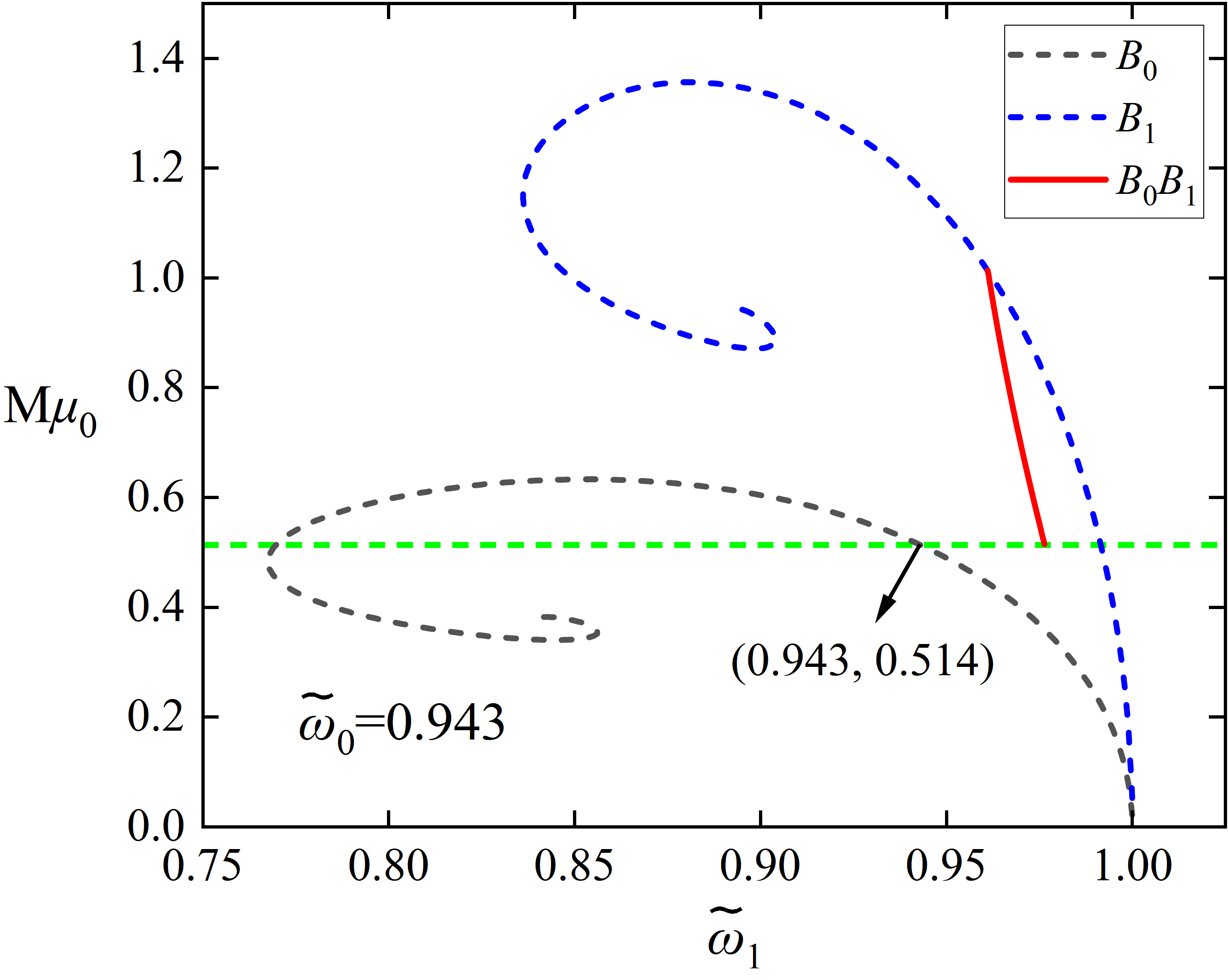}
			\label{fig:nms0.943}
		}	
        \subfigure{  
			\includegraphics[height=.24\textheight, angle =0]{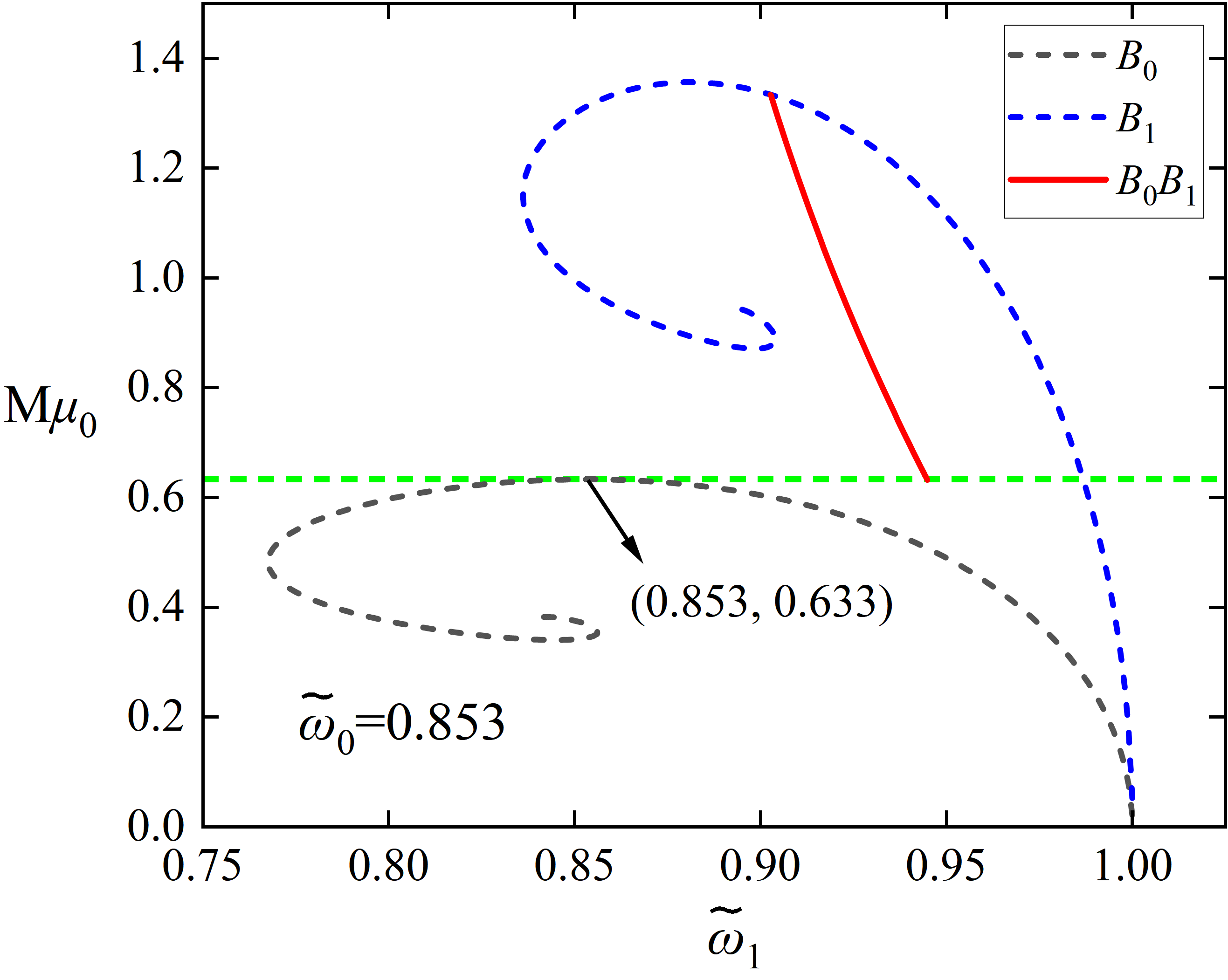}
			\label{fig:nms0.853}
		}
        \subfigure{  
			\includegraphics[height=.24\textheight, angle =0]{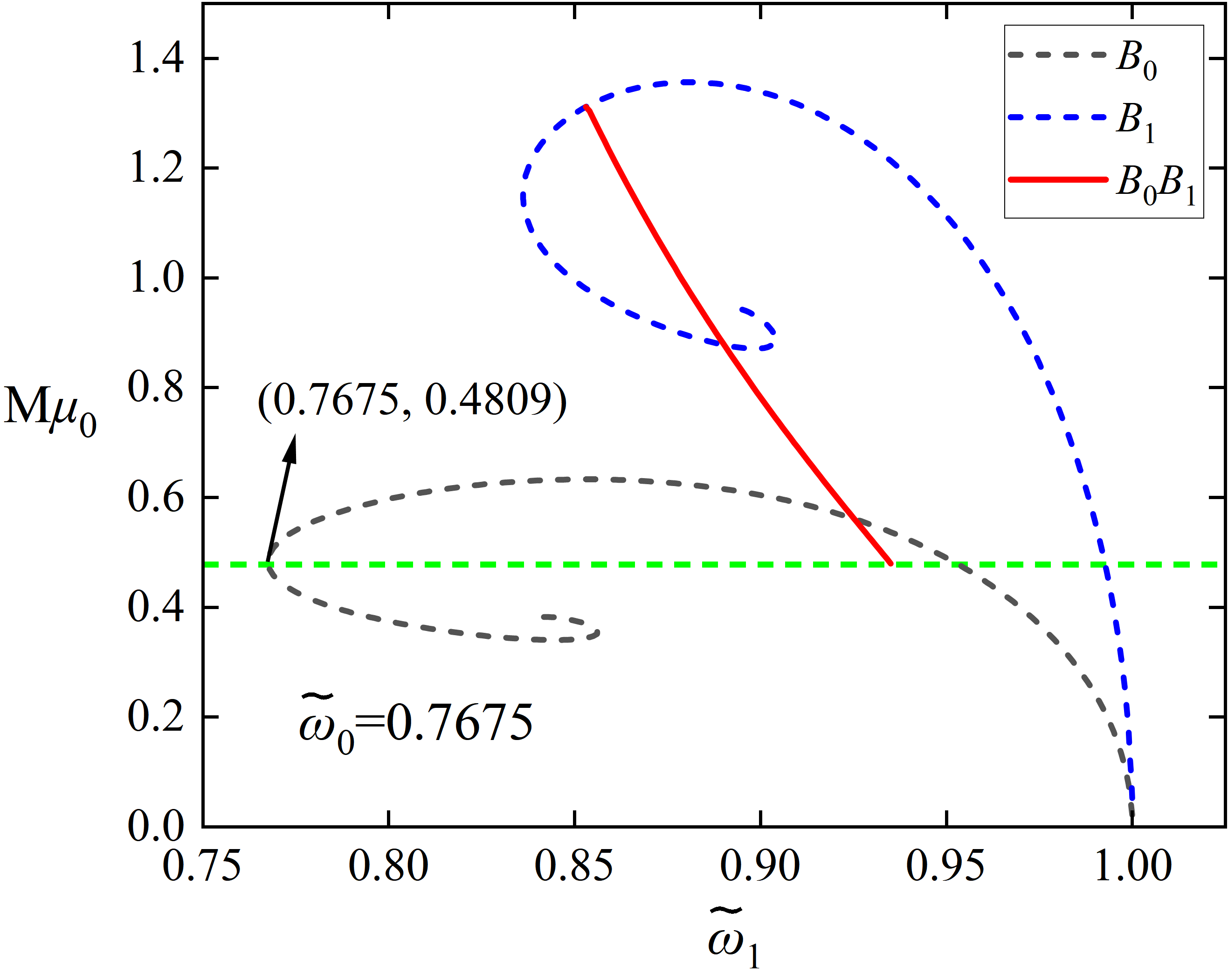}
			\label{fig:nms0.7675}
		}        
  		\end{center}		
		\caption{ The ADM mass $M$ of the MSBSs as a function of $\tilde{\omega}_1$ for $\tilde{\omega}_0=0.985, 0.943, 0.853, 0.7675$. All solutions correspond to $\tilde{\mu}_0=\tilde{\mu}_1=1$. }
	\label{fig:nms}	
		\end{figure}

The relationship between the ADM mass $M$ and the first excited state field frequency $\tilde{\omega}_1$ can be observed from Fig.~\ref{fig:nms}. In the figure, the black and blue dashed lines represent the ground state ($B_0$) and first excited state ($B_1$) boson stars, respectively. The red solid lines correspond to the solutions of MSBSs ($B_0B_1$). The green dashed lines represent the ADM mass of the ground state boson stars for $\tilde{\omega}_0=0.985, 0.943, 0.853, 0.7675$, respectively. From one of the panels, it can be seen that for a fixed $\tilde{\omega}_0$, the ADM mass gradually increases as $\tilde{\omega}_1$ decreases. At the minimum excited state frequency $\tilde{\omega}_1$, the red solid line intersects the blue dashed line, indicating that $B_0B_1$ degenerates into $B_1$. Moreover, it can be observed that the intersection points all lie on the first branch of the excited state boson stars. In addition, at the maximum frequency $\tilde{\omega}_1$, the red solid line intersects the green dashed line, meaning that the MSBSs and the ground state boson stars have the same ADM mass. This is because at that point, the excited state scalar field vanishes, and the MSBSs degenerate into the ground state boson stars. A comparison of these four panels shows that the range of $\tilde{\omega}_1$ over which MSBSs exist increases as $\tilde{\omega}_0$ decreases.

%%%%%%%%%%%%%%%%%%%%%%%%%%%%%%%%%%%%%%%%%%%%%%%%%%%%%%%%%%%		
  %%%%%%%%%%%%%%%%%%%%%%%%%%%%%%%%%%%%%%%%%%%%%%%%%%%%%%%%%%%

  %%%%%%%%%%%%%%%%%%%%%%%%%%%%%%%%%%%%%%%%%%%%%%%%%%%%%%%%%%%		
  %%%%%%%%%%%%%%%%%%%%%%%%%%%%%%%%%%%%%%%%%%%%%%%%%%%%%%%%%%%

\subsubsection{Double-branch}

\begin{figure}[!htbp]
		\begin{center}
		\subfigure{ 
        \includegraphics[height=.28\textheight,width=.32\textheight, angle =0]{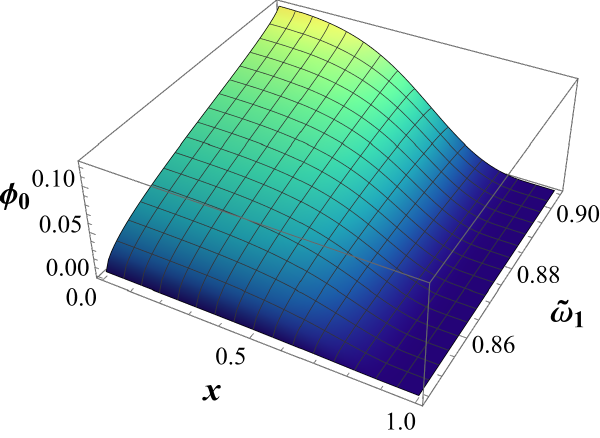}
			\label{fig:nd0.748f01}
		}	 
  		\subfigure{  
			\includegraphics[height=.28\textheight,width=.32\textheight, angle =0]{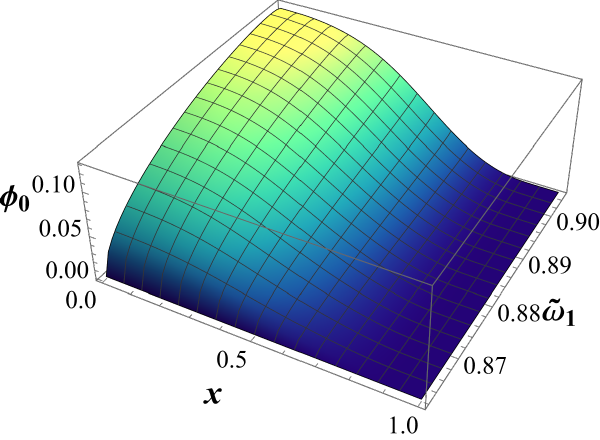}
			\label{fig:nd0.748f02}
		}	
        \subfigure{  
			\includegraphics[height=.28\textheight,width=.32\textheight, angle =0]{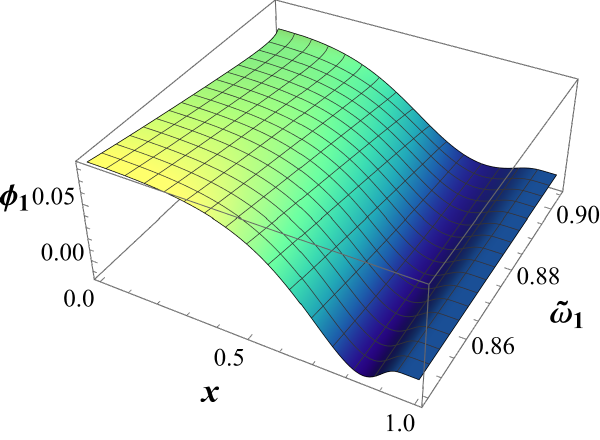}
			\label{fig:nd0.748f11}
		}
        \subfigure{  
			\includegraphics[height=.28\textheight,width=.32\textheight, angle =0]{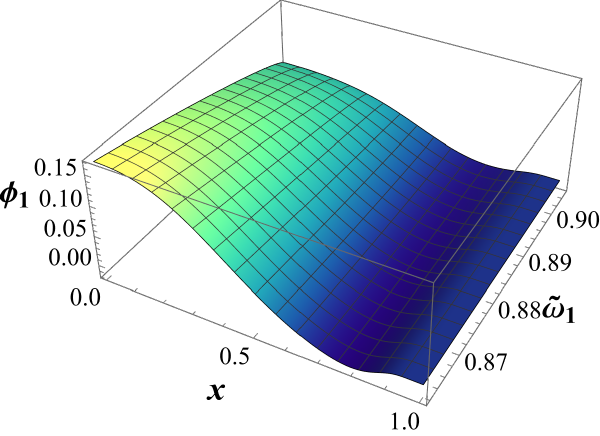}
			\label{fig:nd0.748f12}
		}        
  		\end{center}		
		\caption{ The matter field functions $\phi_0$ and $\phi_1$ on the first (left two panels) and second (right two panels) branches of the MSBSs as functions of $x$ and $\tilde{\omega}_1$ for $\tilde{\omega}_0=0.748$ and $\tilde{\mu}_0=\tilde{\mu}_1=1$. }
	\label{fig:ndf}	
		\end{figure}

The ground state and first excited state scalar field functions for $\tilde{\omega}_0=0.748$ are shown in Fig.~\ref{fig:ndf}. The left panels show the matter field functions $\phi_0$ and $\phi_1$ for the first branch solutions, and the right panels show those for the second branch solutions. From these four panels, it can be seen that for both the first and second branch solutions, as $\tilde{\omega}_1$ increases, the ground state field function $\phi_0$ increases, while the excited state field function $\phi_1$ decreases. Unlike the single-branch case, it can be observed from the lower two panels that $\phi_1$ does not vanish at the maximum frequency $\tilde{\omega}_1$, meaning that the MSBSs cannot degenerate into the ground state boson stars. The upper two panels show that as the frequency $\tilde{\omega}_1$ approaches the minimum, $\phi_0$ of both branches tends to zero, so the MSBSs can degenerate into the first excited state boson stars.

\begin{figure}[!htbp]
		\begin{center}
		\subfigure{ 
        \includegraphics[height=.24\textheight, angle =0]{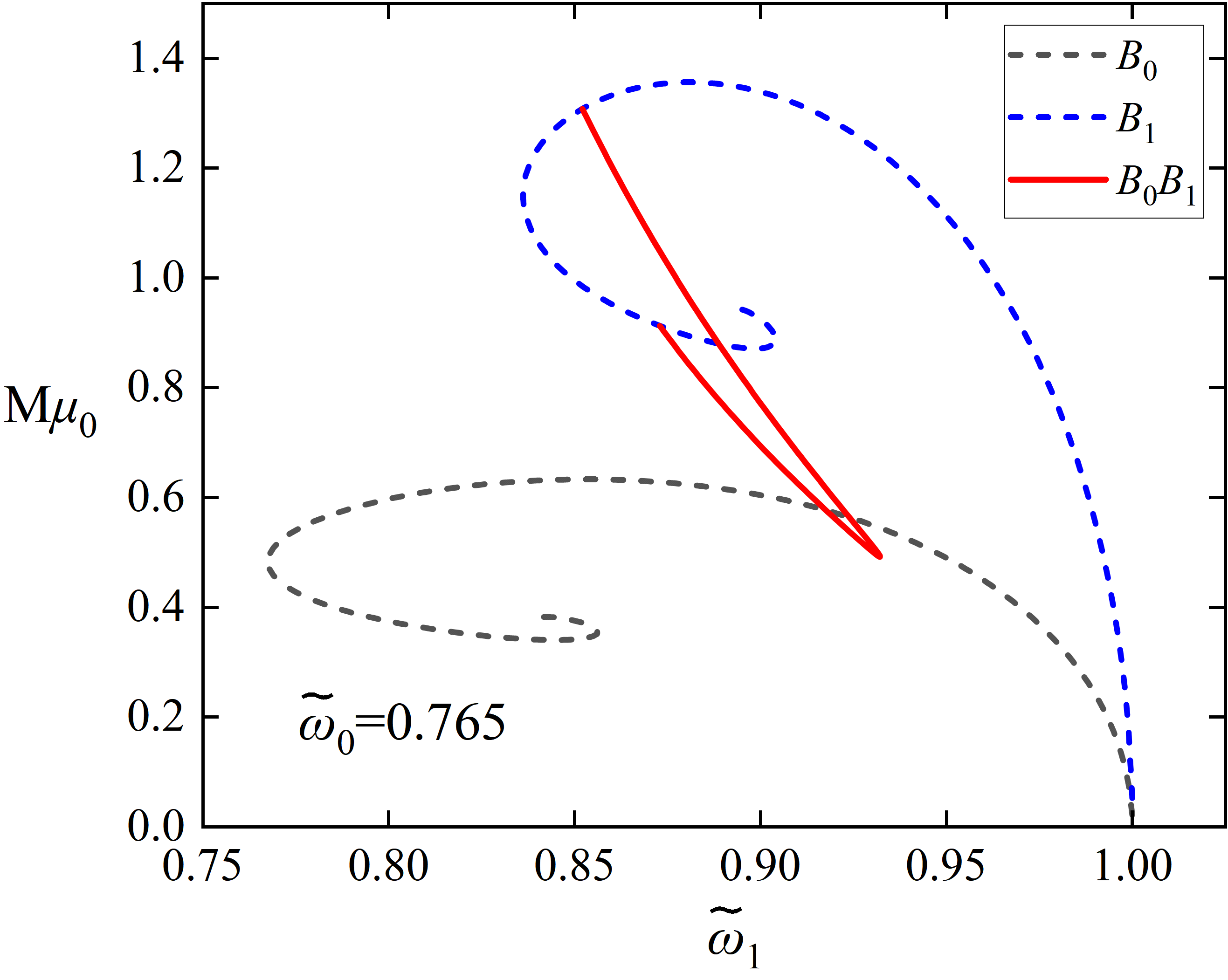}
			\label{fig:nmd0.765}
		}	 
  		\subfigure{  
			\includegraphics[height=.24\textheight, angle =0]{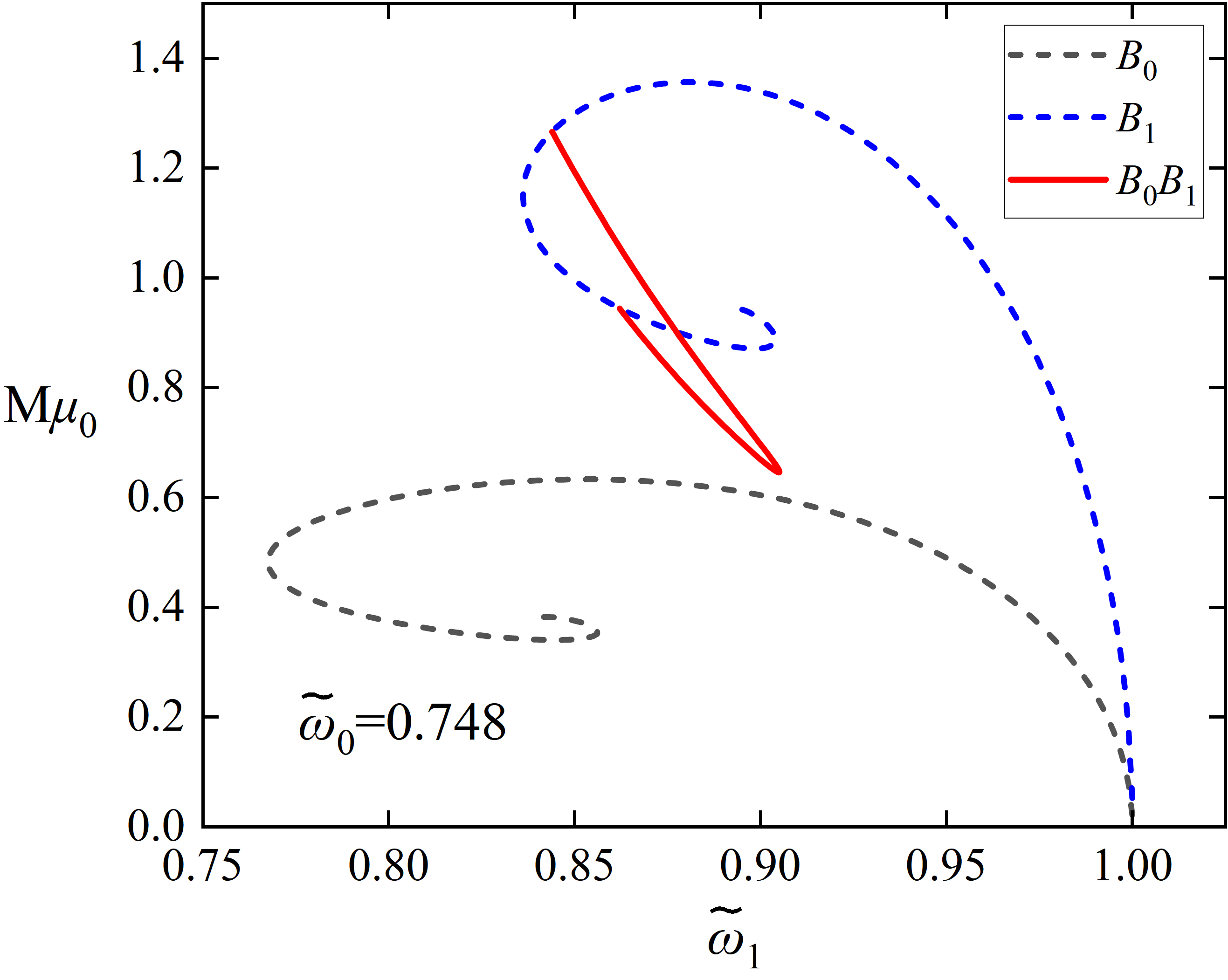}
			\label{fig:nmd0.748}
		}	
        \subfigure{  
			\includegraphics[height=.24\textheight, angle =0]{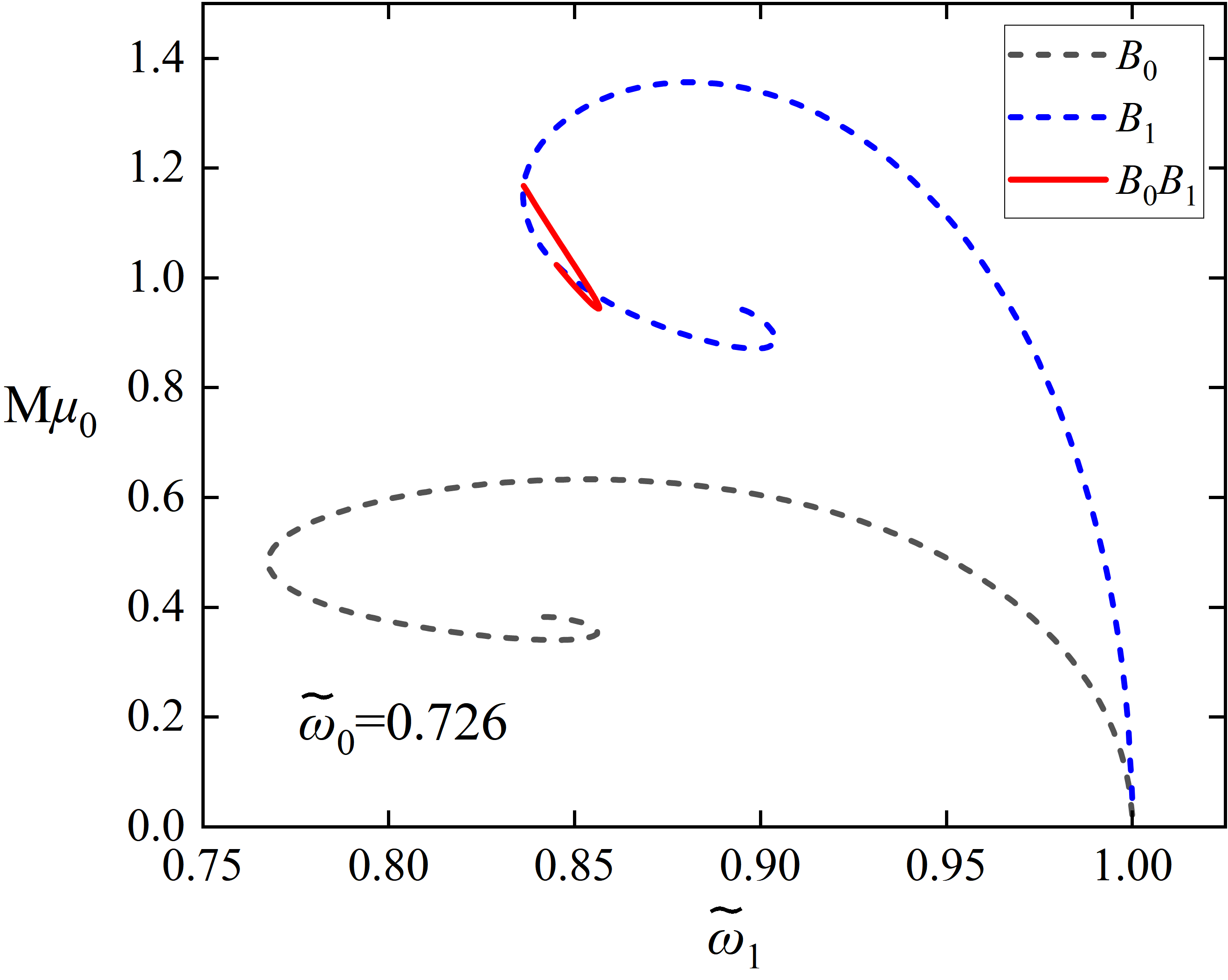}
			\label{fig:nmd0.726}
		}
        \subfigure{  
			\includegraphics[height=.24\textheight, angle =0]{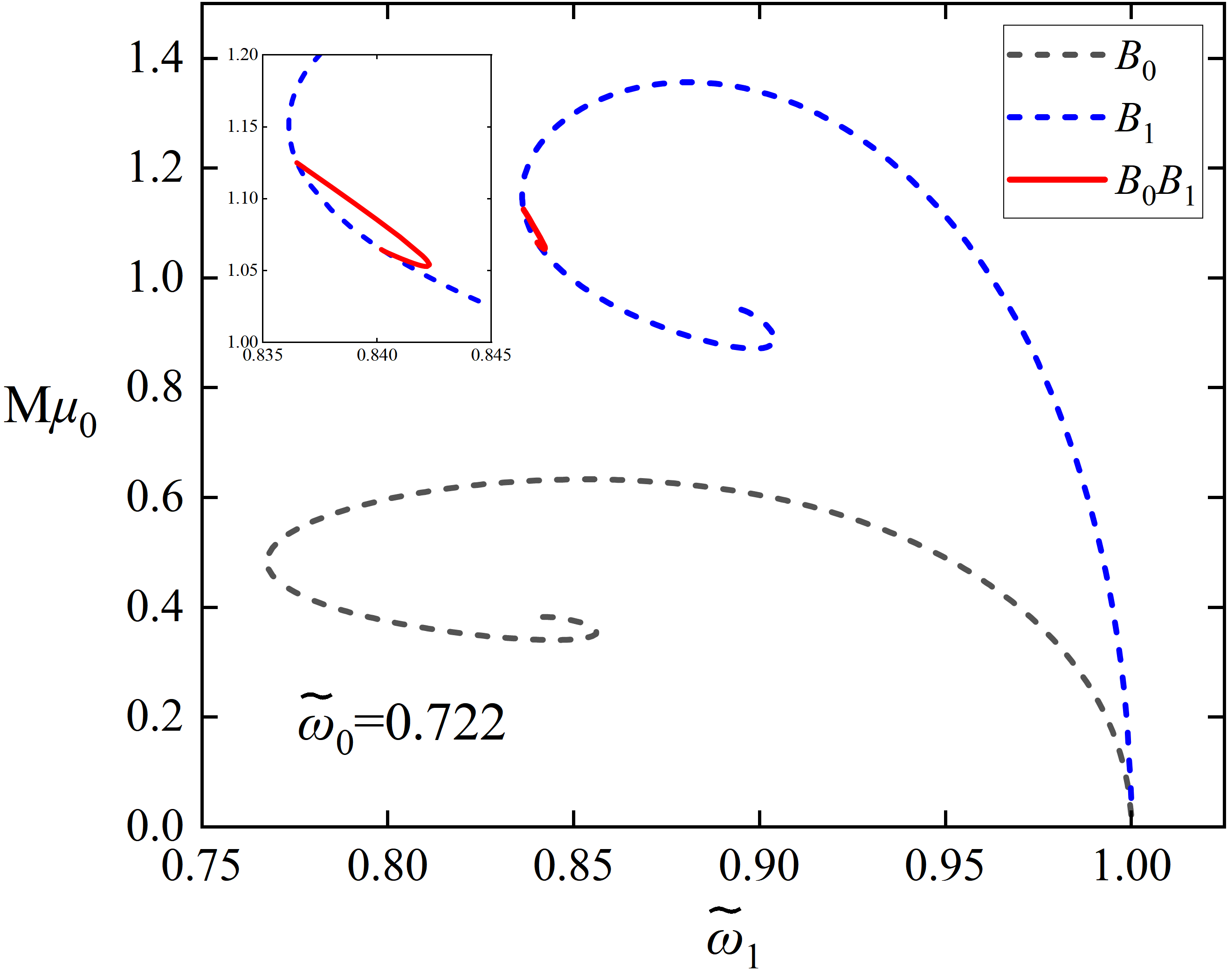}
			\label{fig:nmd0.722}
		}        
  		\end{center}		
		\caption{ The ADM mass $M$ of the MSBSs as a function of $\tilde{\omega}_1$ for $\tilde{\omega}_0=0.765, 0.748, 0.726, 0.722$. All solutions correspond to $\tilde{\mu}_0=\tilde{\mu}_1=1$. }
	\label{fig:nmd}	
		\end{figure}

Fig.~\ref{fig:nmd} shows the relationship between the ADM mass $M$ and the first excited state field frequency $\tilde{\omega}_1$. The red solid lines represent the double-branch solutions, with the longer curves corresponding to the first branch and the shorter ones to the second branch. From any panel, it can be seen that at the minimum value of $\tilde{\omega}_1$, the first branch intersects the blue dashed line, indicating that the MSBSs degenerate into the first excited state boson stars. However, at the maximum value of $\tilde{\omega}_1$, the MSBSs do not degenerate into the ground state boson stars but instead turn into the second branch. The second branch eventually degenerates into the first excited state boson stars as the frequency $\tilde{\omega}_1$ decreases. Comparing these four panels, it is evident that the range of $\tilde{\omega}_1$ for the existence of MSBSs decreases as $\tilde{\omega}_0$ decreases.
  %%%%%%%%%%%%%%%%%%%%%%%%%%%%%%%%%%%%%%%%%%%%%%%%%%%%%%%%%%%		
  %%%%%%%%%%%%%%%%%%%%%%%%%%%%%%%%%%%%%%%%%%%%%%%%%%%%%%%%%%%

  %%%%%%%%%%%%%%%%%%%%%%%%%%%%%%%%%%%%%%%%%%%%%%%%%%%%%%%%%%%		
 %%%%%%%%%%%%%%%%%%%%%%%%%%%%%%%%%%%%%%%%%%%%%%%%%%%%%%%%%%%
\subsection{Binding energy}

We analyze the stability of the MSBSs from the perspective of binding energy. The binding energy of the MSBSs is given by
\begin{equation}\label{eqube}
     E_B = M-\mu_0 Q_0-\mu_1 Q_1\, , 
\end{equation}
where $M$ is the ADM mass of the MSBSs, and $Q_0$ and $Q_1$ are the Noether charges of the ground state and first excited state scalar fields, respectively.

\begin{figure}[!htbp]
		\begin{center}
		\subfigure{ 
        \includegraphics[height=.24\textheight, angle =0]{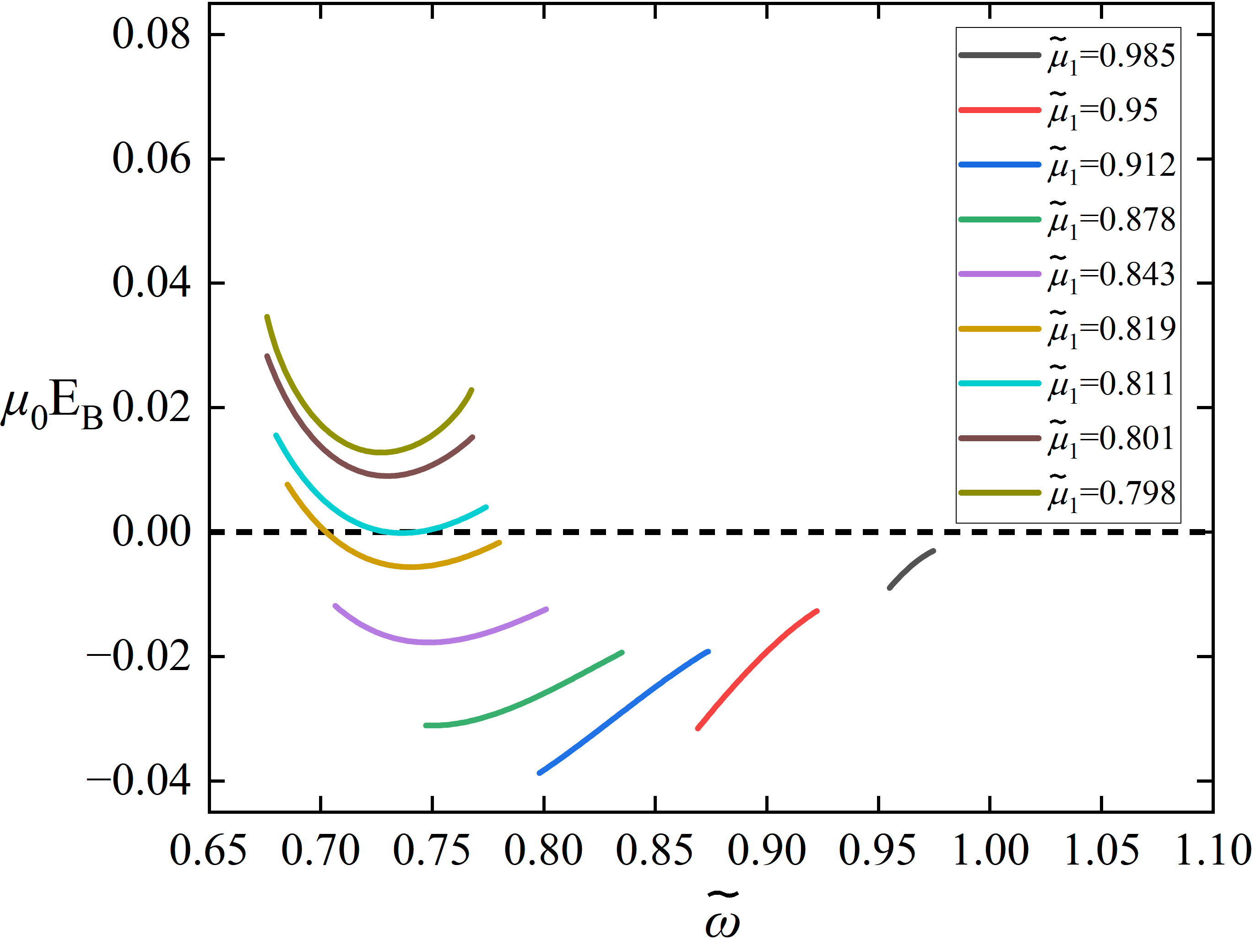}
			\label{fig:besall}
		}	 
  		\subfigure{  
			\includegraphics[height=.24\textheight, angle =0]{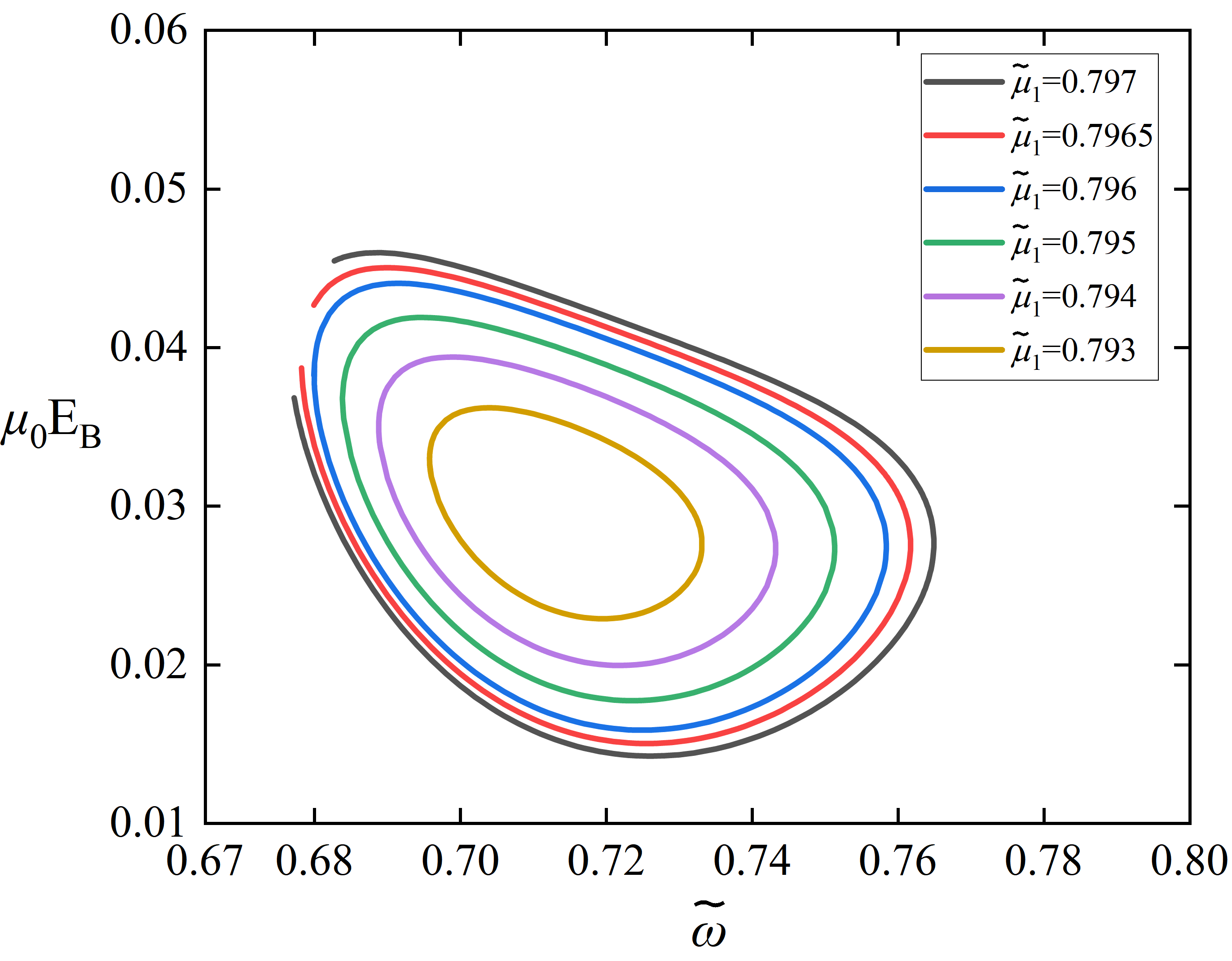}
			\label{fig:bedall}
		}	        
  		\end{center}		
		\caption{ Left: The binding energy $E_B$ of the single-branch solutions as a function of the synchronized frequency $\tilde{\omega}$ for several values of $\tilde{\mu}_1$. Right: Same as left panel for the double-branch solutions. All solutions correspond to $\tilde{\mu}_0=1$. }
	\label{fig:be}	
		\end{figure}

We first analyze the stability of the MSBSs under the synchronized frequency condition. In general, the stability of boson stars can be preliminarily assessed by the binding energy: stable solutions have negative binding energy, while positive binding energy indicates instability. As shown in Fig.~\ref{fig:be}, the left panel shows the relationship between the binding energy $E_B$ of the single-branch solutions and synchronized frequency $\tilde{\omega}$, and the right panel shows the corresponding relationship for the double-branch solutions. From the left panel, it can be seen that for $0.828 \le \tilde{\mu}_1<1$, the binding energy of the whole branch satisfies $E_B<0$. As $\tilde{\mu}_1$ decreases, stable and unstable solutions appear within the same branch. When $\tilde{\mu}_1$ decreases to the range $0.7976 \le \tilde{\mu}_1<0.811$, the solutions in the entire branch have $E_B>0$, indicating that all solutions in this branch are unstable. Thus, MSBSs are stable when $\tilde{\mu}_1$ is large enough, i.e., when the two field masses are close. The right panel shows that the $E_B$ of the double-branch solutions is always positive, which means these solutions are unstable. When $\tilde{\mu}_1\le 0.796$, the binding energy curve exhibits a loop-like structure, and the size of the loop decreases as $\tilde{\mu}_1$ decreases.

 %%%%%%%%%%%%%%%%%%%%%%%%%%%%%%%%%%%%%%%%%%%%%%%%%%%%%%%%%%%		
 %%%%%%%%%%%%%%%%%%%%%%%%%%%%%%%%%%%%%%%%%%%%%%%%%%%%%%%%%%%

\begin{figure}[!htbp]
		\begin{center}
		\subfigure{ 
        \includegraphics[height=.24\textheight, angle =0]{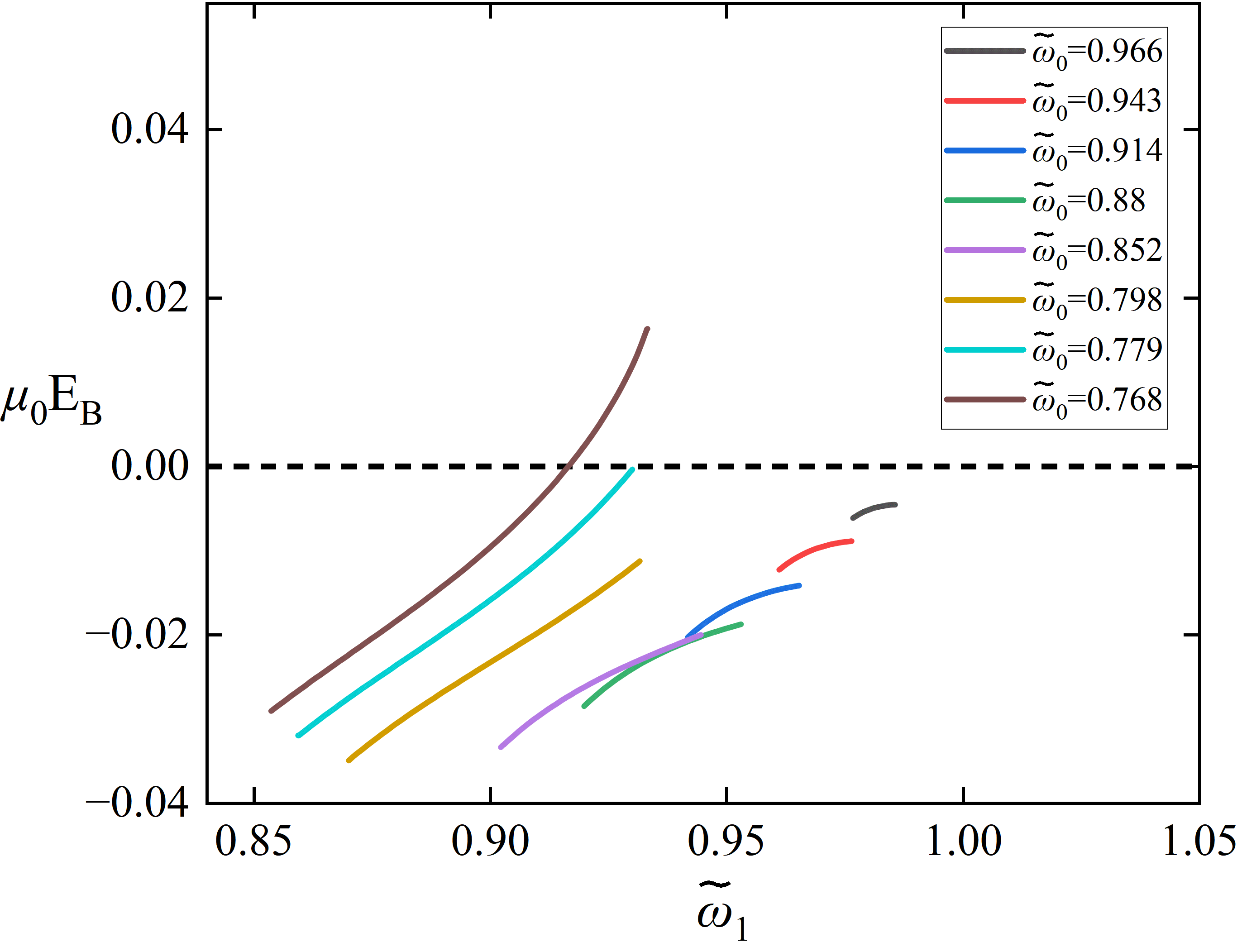}
			\label{fig:bensall}
		}	 
  		\subfigure{  
			\includegraphics[height=.24\textheight, angle =0]{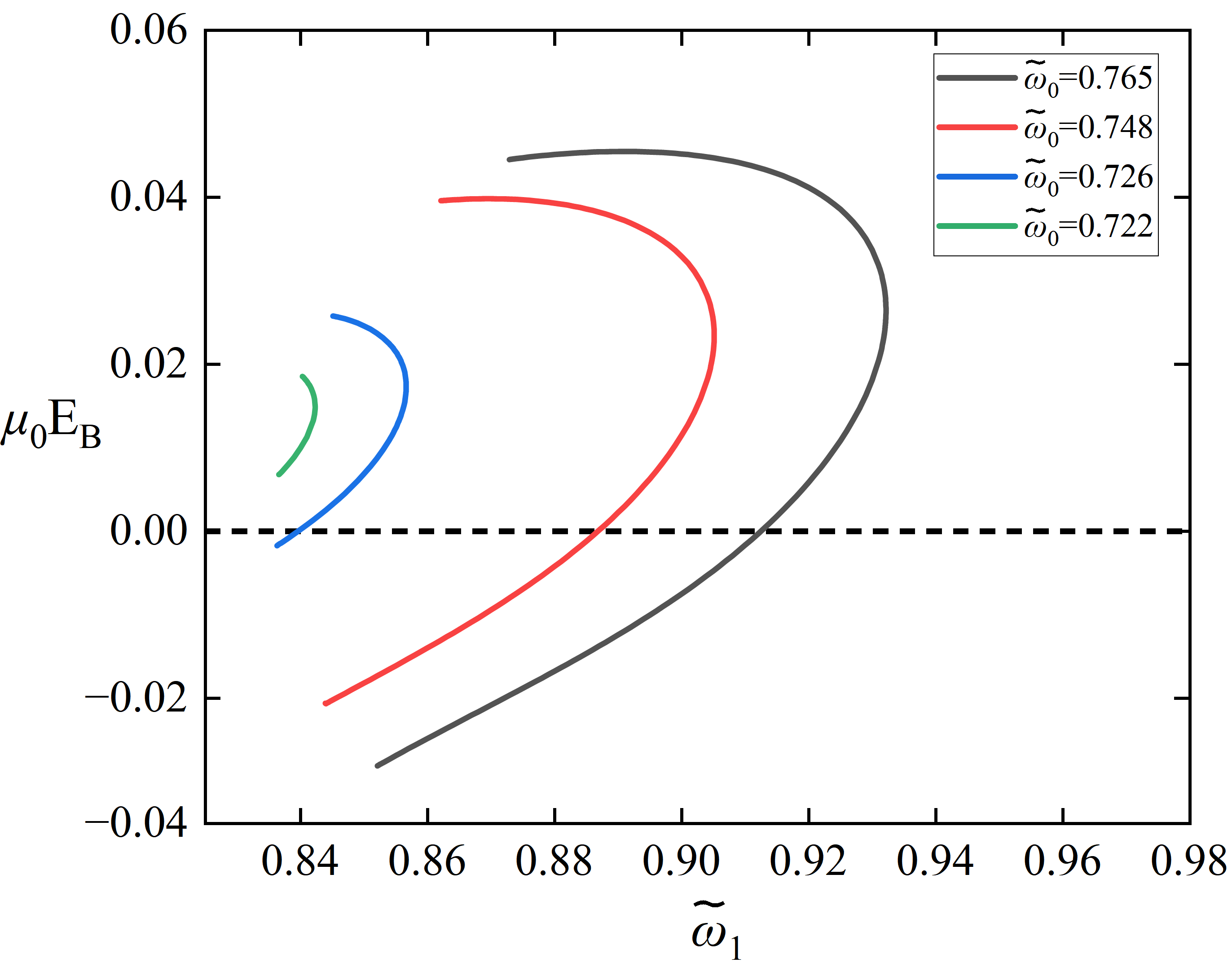}
			\label{fig:bendall}
		}	        
  		\end{center}		
		\caption{ Left: The binding energy $E_B$ of the single-branch solutions as a function of the frequency $\tilde{\omega}_1$ for several values of $\tilde{\omega}_0$. Right: Same as left panel for the double-branch solutions. All solutions correspond to $\tilde{\mu}_0=\tilde{\mu}_1=1$. }
	\label{fig:ben}	
		\end{figure}

Fig.~\ref{fig:ben} shows the binding energy $E_B$ under the nonsynchronized frequency condition. The left panel displays $E_B$ for single-branch solutions. For a fixed $\tilde{\omega}_0$, the binding energy $E_B$ increases as $\tilde{\omega}_1$ increases. $E_B$ is always negative for $0.779\le\tilde{\omega}_0<1$, indicating stability. However, when $\tilde{\omega}_0<0.779$, $E_B$ changes from negative to positive as $\tilde{\omega}_1$ increases, and the solutions become unstable. The right panel shows $E_B$ for double-branch solutions. The lower and upper parts of the loop-shaped curve correspond to the first and second branches, respectively. It can be seen from the figure that the $E_B$ of the second branch solutions is always positive, indicating that these solutions are unstable. For the first branch solutions, when $\tilde{\omega}_0$ is relatively large (e.g., greater than $0.726$), as $\tilde{\omega}_1$ increases, the solutions transition from stable to unstable. Moreover, as $\tilde{\omega}_0$ decreases, the range of $\tilde{\omega}_1$ for which stable solutions exist also decreases. When $\tilde{\omega}_0$ is sufficiently small (e.g., equal to $0.722$), there are no stable solutions in the first branch either.

 %%%%%%%%%%%%%%%%%%%%%%%%%%%%%%%%%%%%%%%%%%%%%%%%%%%%%%%%%%%		
 %%%%%%%%%%%%%%%%%%%%%%%%%%%%%%%%%%%%%%%%%%%%%%%%%%%%%%%%%%%

\section{TIDAL LOVE NUMBERS OF MULTI-STATE BOSON STARS}\label{Sec5}

In this section, we discuss the tidal Love numbers of spherically symmetric MSBSs. We take a location far outside the effective radius of the MSBSs as $R_\text{ext}$. In the numerical calculation of the electric tidal Love numbers, in order to eliminate drastic variations at the boundaries and improve the accuracy and stability of the numerical solution, we perform the following transformations on $H_0$, $\psi_0$ and $\psi_1$~\cite{Sennett:2017etc}
\begin{equation}\label{equHr2}
	\tilde{H}_0(r) = H_0 r^{-2}, \qquad \tilde{\psi}_0(r) = \psi_0 r^{-3}, \qquad \tilde{\psi}_1(r) = \psi_1 r^{-3}\,.
\end{equation}

In order to solve Eqs. (\ref{equH}), (\ref{equpsi0}), and (\ref{equpsi1}), we need to impose appropriate boundary conditions, which are given as follows:
\begin{equation}\label{equHbon}
\begin{split}
	\tilde{H}_0(0) &= \tilde{H}_0^{(c)},\quad \tilde{H}_0'(0) = 0,\\
    \tilde{\psi}'_0(0) &= 0,\quad \tilde{\psi}_0(\infty) = 0,\\
    \tilde{\psi}'_1(0) &= 0,\quad \tilde{\psi}_1(\infty) = 0,
\end{split}
\end{equation}
where $\tilde{H}_0^{(c)}$ is an arbitrary non-zero constant, and we take $\tilde{H}_0^{(c)}=1$.

As in the electric perturbation case, we need to perform the following transformation on $h_0$
\begin{equation}\label{equh3}
	\tilde{h}_0(r) = h_0 \, r^{3}\,.
\end{equation}
In order to solve Eq. (\ref{equh0}), the following boundary conditions need to be imposed:
\begin{equation}\label{equhbon}
	\tilde{h}_0(0) = \tilde{h}_0^{(c)}, \qquad \tilde{h}_0'(0) = 0\,,
\end{equation}
where $\tilde{h}_0^{(c)}$ is an arbitrary non-zero constant, and we also choose $\tilde{h}_0^{(c)}=1$.

By calculating the binding energy of the background solutions, we find that the binding energy of all configurations of the synchronized frequency double-branch solutions and the second branch of the nonsynchronized frequency double-branch solutions is positive. This implies that these solution configurations are unstable. Therefore, we do not calculate the tidal Love numbers for these solutions, and only calculate the tidal Love numbers for branches that contain negative binding energy.

\begin{figure}[!htbp]
		\begin{center}
		\subfigure{ 
        \includegraphics[height=.24\textheight, angle =0]{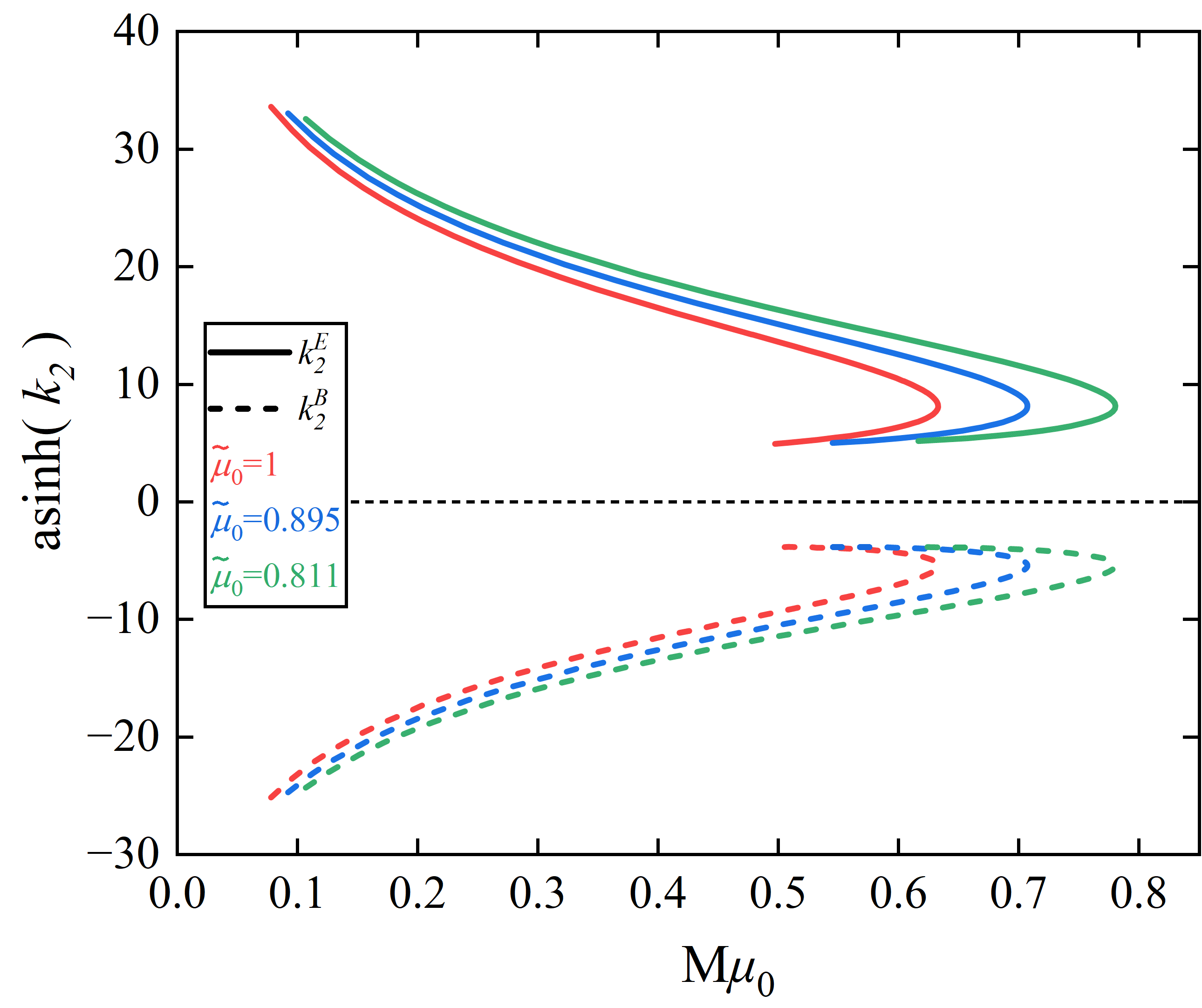}
		}	 
  		\subfigure{  
			\includegraphics[height=.24\textheight, angle =0]{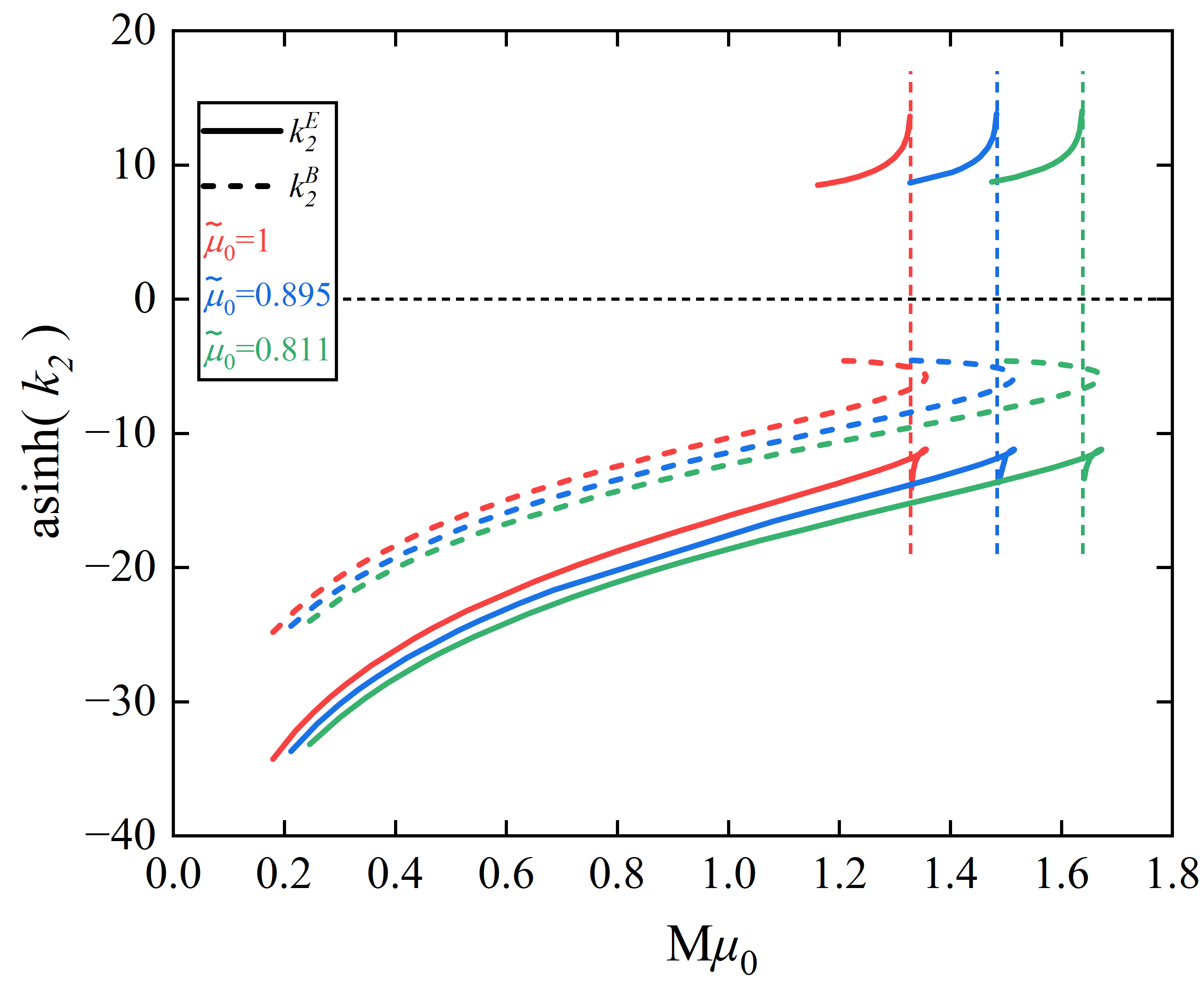}
		}	       
  		\end{center}		
		\caption{ The electric (solid line) and magnetic (dashed line) tidal Love numbers of the ground state (left) and first excited state (right) boson stars for different field masses $\tilde{\mu}_0$ and $\tilde{\mu}_1$. }
	\label{fig:tidal01}	
		\end{figure}

Before presenting the tidal Love numbers of MSBSs, we first show the tidal Love numbers of single-state boson stars. The Love numbers of the stable solutions of ground state boson stars, ranging from the maximum frequency to the maximum mass~\cite{Lee:1988av,Gleiser:1988rq}, have already been calculated~\cite{Cardoso:2017cfl,Mendes:2016vdr,Sennett:2017etc}. Here, we supplement the Love numbers for the remaining part of the first branch of ground state boson stars as well as those of the first excited state boson stars. When plotting all the tidal Love numbers, for ease of presentation, we plot $\operatorname{asinh}(k)$, where $\operatorname{asinh}(k) = \ln(k + \sqrt{k^2+1})$. As shown in Fig. \ref{fig:tidal01}, the left and right panels show the tidal Love numbers of the ground state and the first excited state boson stars, respectively. In the figure, only the first branch solutions are shown. Except for the electric Love numbers of the first excited state boson stars, the other three Love numbers change continuously with the mass $M$. From the minimum to the maximum mass, the absolute values of the Love numbers decrease with increasing mass, implying that the boson stars become less deformable. For fixed $M$, the absolute values of the Love numbers increase as $\tilde{\mu}_0$ and $\tilde{\mu}_1$ increase. The electric tidal Love numbers are positive, while the magnetic tidal Love numbers are negative. The positive (negative) sign indicates that the response of the object to the external tidal field is positive (negative) feedback. The electric tidal Love numbers of the first excited state boson stars are significantly different from those of the other three cases. For $\tilde{\mu}_1=1, 0.895$ and $0.811$, starting from the minimum mass, $k_2^E$ is initially negative and its absolute value decreases as $M$ increases. After reaching the maximum mass, the absolute value of $k_2^E$ increases until the masses reach approximately $1.3287$, $1.4850$, and $1.6393$, respectively, at which point it jumps from negative to positive and then gradually decreases. Near these specific masses, a small change in the mass of boson stars leads to a dramatic change in the Love numbers, accompanied by a sign change. We refer to this phenomenon as a ``peak''.

%%%%%%%%%%%%%%%%%%%%%%%%%%%%%%%%%%%%%%%%%%%%%%%%%%%%%%%%%%%		
  %%%%%%%%%%%%%%%%%%%%%%%%%%%%%%%%%%%%%%%%%%%%%%%%%%%%%%%%%%%
\subsection{Case \uppercase\expandafter{\romannumeral1}: $\omega_0 = \omega_1$, single-branch}

\begin{figure}[!htbp]
		\begin{center}
		\subfigure{ 
        \includegraphics[height=.24\textheight, angle =0]{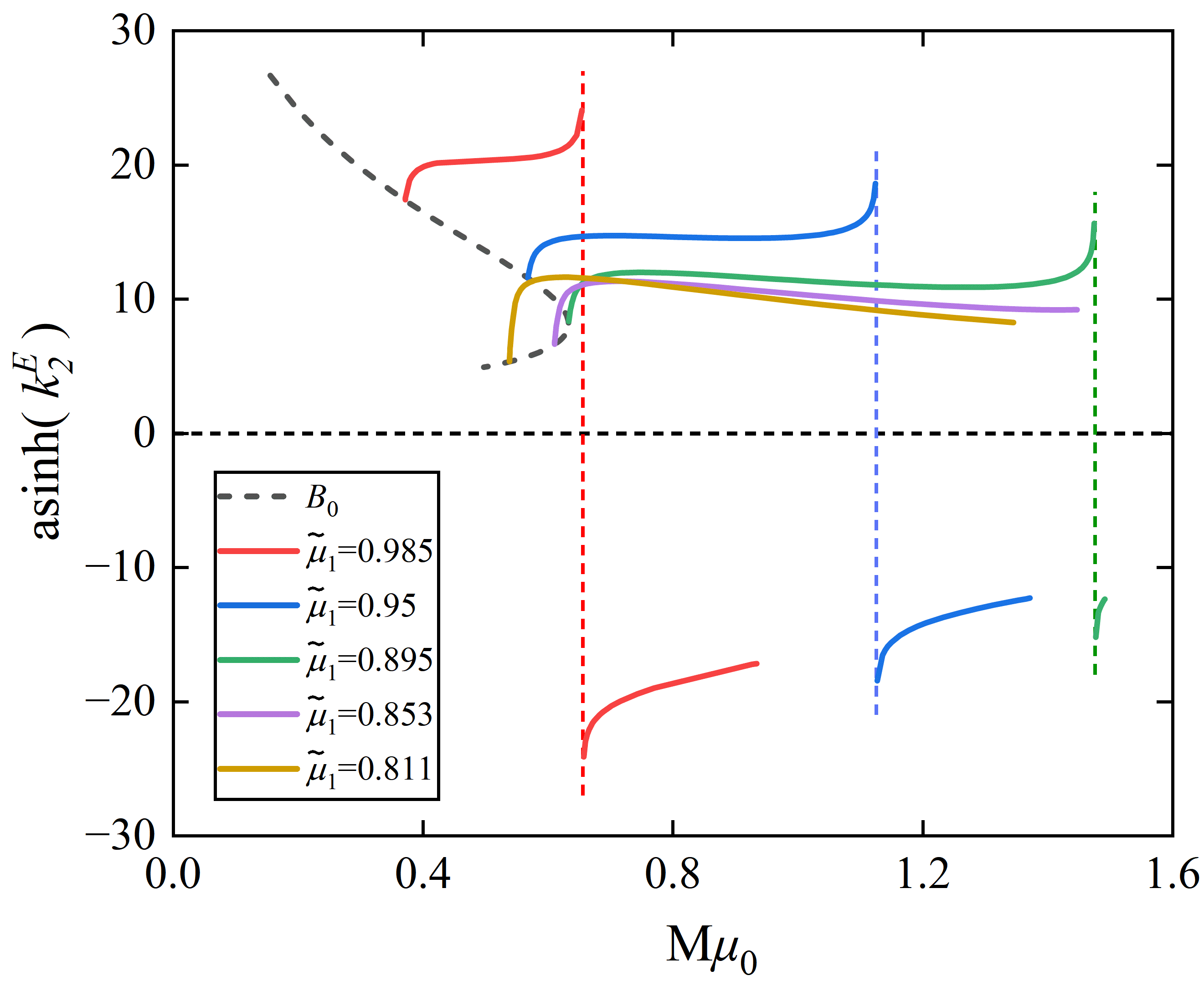}
			\label{fig:stidal}
		}	 
  		\subfigure{  
			\includegraphics[height=.24\textheight, angle =0]{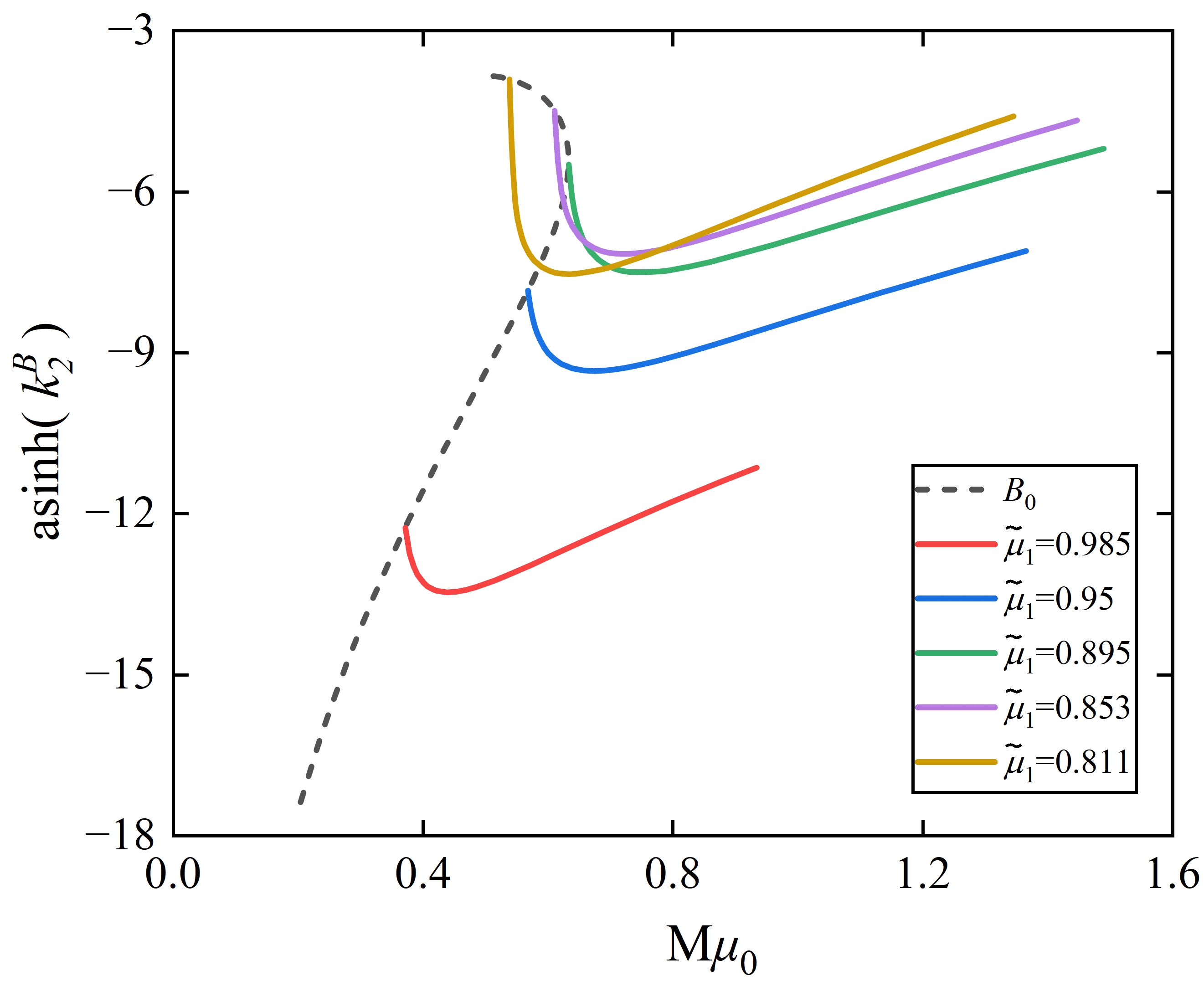}
			\label{fig:stidalo}
		}	        
  		\end{center}		
		\caption{ Left: The electric tidal Love numbers as a function of the mass of the MSBSs in the case of synchronized frequency single-branch solutions. Right: Same as the left panel, but for the magnetic tidal Love numbers. }
	\label{fig:stidalall}	
		\end{figure}

Next, we present the tidal Love numbers of MSBSs. First, we discuss the tidal Love numbers of the synchronized frequency single-branch solutions. As shown in Fig. \ref{fig:stidalall}, the left panel presents the electric tidal Love numbers $k_2^E$ for different values of the excited state field mass $\tilde{\mu}_1$. The right panel shows the magnetic tidal Love numbers $k_2^B$. In both panels, the black dashed line represents the tidal Love numbers of the ground state boson stars. It can be seen from the red solid line in the left panel that when $\tilde{\mu}_1=0.985$, as the mass $M$ increases, the Love numbers $k_2^E$ are initially positive and increase. When the mass $M$ reaches approximately $0.6558$, the Love numbers suddenly jump to negative values, and the absolute values decrease with increasing $M$. The blue and green solid lines correspond to the $k_2^E$ for $\tilde{\mu}_1=0.95$ and $\tilde{\mu}_1=0.895$, which are slightly different from the case $\tilde{\mu}_1=0.985$. As the mass $M$ increases, $k_2^E$ start positive, first increase, then decrease, and increase again, until the values suddenly jump to negative at masses of approximately $1.1257$ and $1.4757$, respectively. For these negative values, the absolute values decrease as $M$ increases. From Fig. \ref{fig:ms}, it can be seen that the mass $M$ decreases monotonically as the synchronized frequency $\tilde{\omega}$ increases. The behavior of these three solid lines shows that, for a fixed $\tilde{\mu}_1$, as $\tilde{\omega}$ decreases and approaches a certain value, the deformation response of the MSBSs under even-parity tidal perturbations becomes maximal and undergoes a sudden transition from positive to negative feedback. In these three cases, a ``peak'' similar to that of the first excited state boson stars appears. It can also be observed from the figure that when $\tilde{\mu}_1$ decreases to a certain value (calculated to be $\tilde{\mu}_1=0.891$), this sudden transition no longer occurs, and the $k_2^E$ are all positive. For $\tilde{\mu}_1=0.811$, $k_2^E$ are all positive as the mass increases, initially increasing and then decreasing. The right panel shows the magnetic tidal Love numbers $k_2^B$ under the same conditions. It can be seen that all magnetic tidal Love numbers are negative, which means that the induced current multipole moments of the MSBSs are opposite to the magnetic multipole moments of the external tidal field. For a fixed value of $\tilde{\mu}_1$, the absolute values of $k_2^B$ first increase and then decrease as $M$ increases. The left endpoints of all solid lines in the figure intersect the black dashed line, indicating once again that the MSBSs can degenerate into the ground state boson stars. 

%%%%%%%%%%%%%%%%%%%%%%%%%%%%%%%%%%%%%%%%%%%%%%%%%%%%%%%%%%%		
 %%%%%%%%%%%%%%%%%%%%%%%%%%%%%%%%%%%%%%%%%%%%%%%%%%%%%%%%%%%
\subsection{Case \uppercase\expandafter{\romannumeral2}: $\omega_0 \ne \omega_1$, single-branch}

\begin{figure}[!htbp]
		\begin{center}
		\subfigure{ 
        \includegraphics[height=.24\textheight, angle =0]{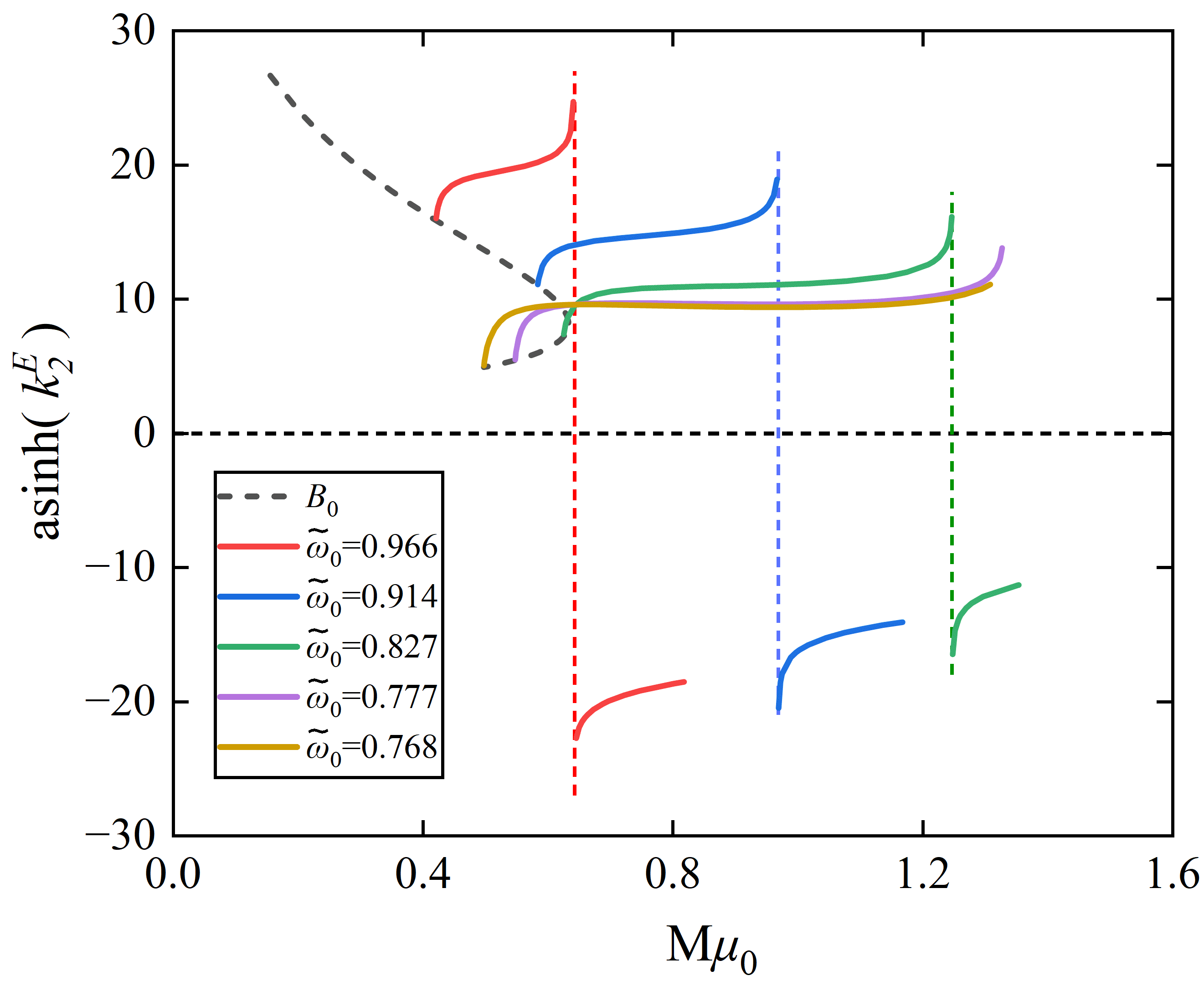}
			\label{fig:nstidal}
		}	 
  		\subfigure{  
			\includegraphics[height=.24\textheight, angle =0]{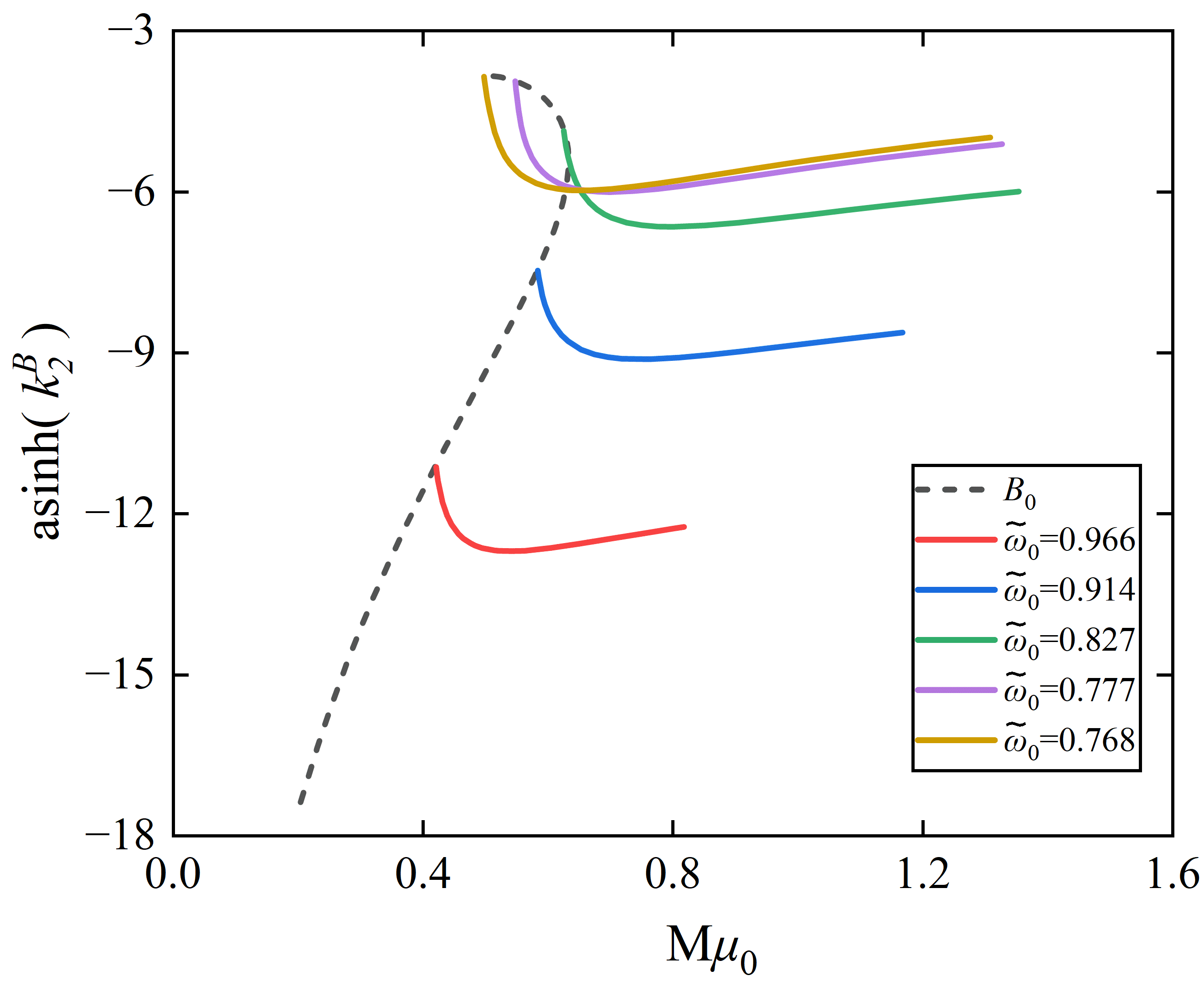}
			\label{fig:nstidalo}
		}	        
  		\end{center}		
		\caption{ Left: The electric tidal Love numbers as a function of the mass of the MSBSs in the case of nonsynchronized frequency single-branch solutions. Right: Same as the left panel, but for the magnetic tidal Love numbers. }
	\label{fig:nstidalall}	
		\end{figure}

Next, we analyze the tidal Love numbers of the single-branch solutions under the nonsynchronized frequency condition. As shown in Fig. \ref{fig:nstidalall}, the left panel presents the electric tidal Love numbers $k_2^E$ for different values of the ground state field frequency $\tilde{\omega}_0$, and the right panel shows the magnetic tidal Love numbers $k_2^B$ under the same conditions. The black dashed line in each panel represents the tidal Love numbers of the ground state boson stars. In the left panel, the red, blue, and green solid lines correspond to the $k_2^E$ for $\tilde{\omega}_0=0.996, 0.914, 0.827$, respectively. It is evident that these three curves also exhibit the ``peak''. These Love numbers are initially positive and increase with increasing mass $M$, until the values suddenly jump to negative when the masses reach approximately $0.6430, 0.9680,$ and $1.2473$, respectively. After the jump, the absolute values of the negative Love numbers decrease as the mass increases. Similar to the synchronized frequency single-branch case, for a fixed $\tilde{\omega}_0$, when $\tilde{\omega}_1$ approaches a specific value, the MSBSs become highly deformable under an external tidal field, and a slight decrease of $\tilde{\omega}_1$ causes the deformation response to transition from positive to negative feedback. Calculations show that when $\tilde{\omega}_0 \le 0.777$, $k_2^E$ do not become negative. Moreover, it can be seen from the figure that as the mass $M$ increases, $k_2^E$ first increase, then decrease, and finally increase again. The right panel shows the magnetic tidal Love numbers $k_2^B$ under the same conditions. For a fixed $\tilde{\omega}_0$, $k_2^B$ are all negative, and the absolute values first decrease and then increase with increasing mass $M$. In both panels, the left endpoints of all solid lines intersect the black dashed line, indicating that the MSBSs degenerate into the ground state boson stars.

%%%%%%%%%%%%%%%%%%%%%%%%%%%%%%%%%%%%%%%%%%%%%%%%%%%%%%%%%%%		
 %%%%%%%%%%%%%%%%%%%%%%%%%%%%%%%%%%%%%%%%%%%%%%%%%%%%%%%%%%%
\subsection{Case \uppercase\expandafter{\romannumeral3}: $\omega_0 \ne \omega_1$, double-branch}

\begin{figure}[!htbp]
		\begin{center}
		\subfigure{ 
        \includegraphics[height=.24\textheight, angle =0]{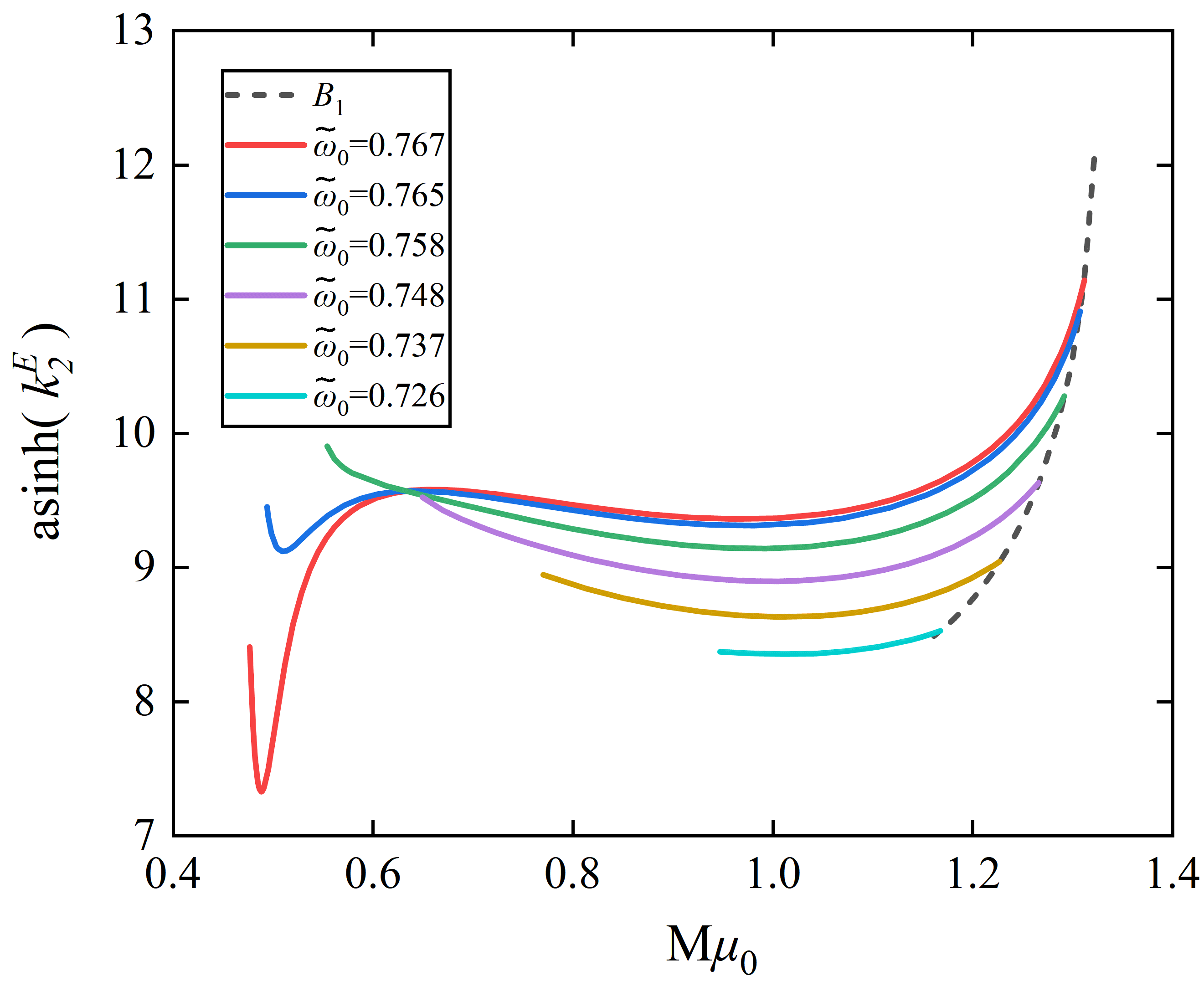}
			\label{fig:ndtidal}
		}	 
  		\subfigure{  
			\includegraphics[height=.24\textheight, angle =0]{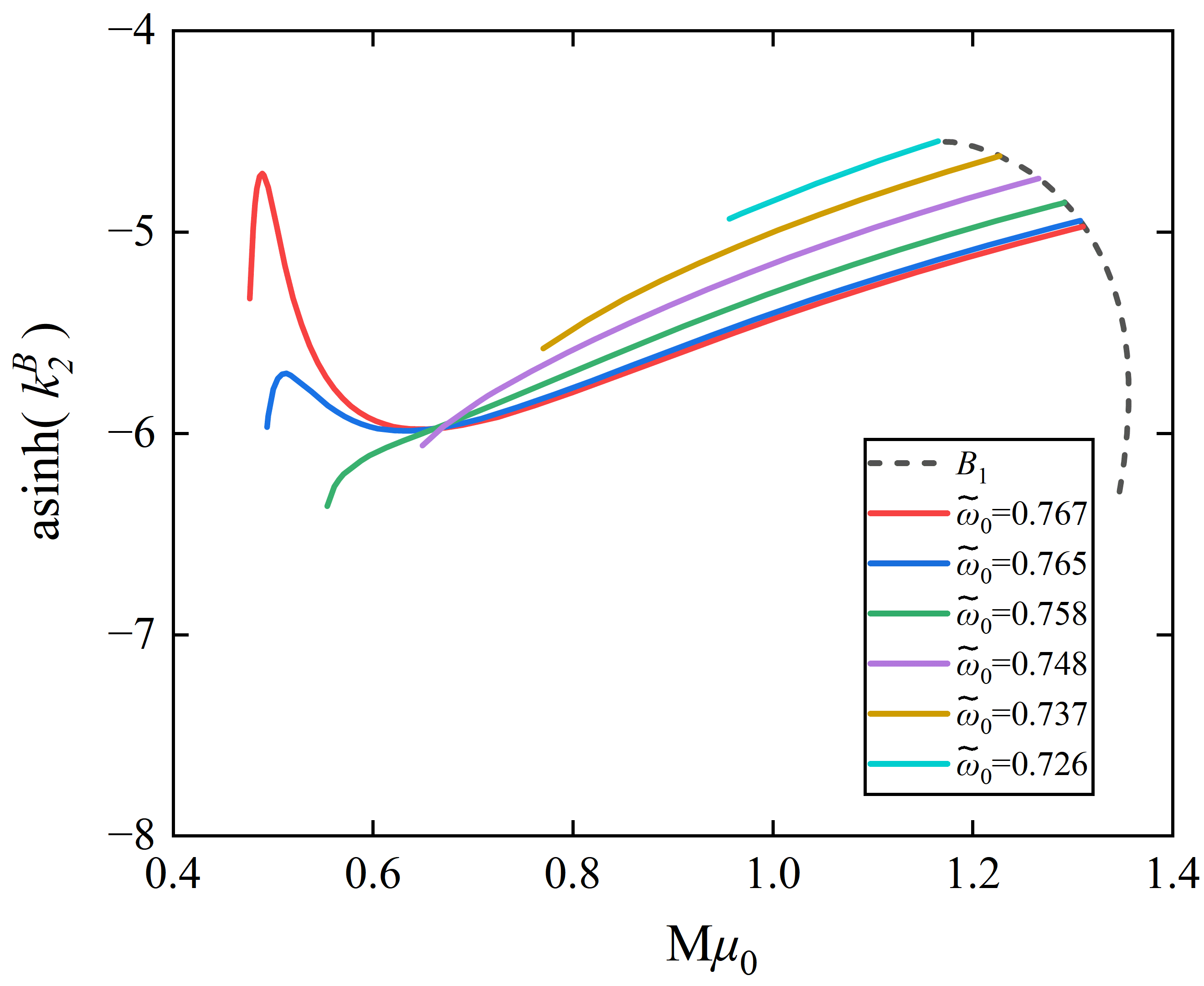}
			\label{fig:ndtidalo}
		}	        
  		\end{center}		
		\caption{ Left: The electric tidal Love numbers as a function of the mass of the MSBSs in the case of nonsynchronized frequency double-branch solutions. Right: Same as the left panel, but for the magnetic tidal Love numbers. }
	\label{fig:ndtidalall}	
		\end{figure}

Finally, we calculate the tidal Love numbers of the double-branch solutions under the nonsynchronized frequency condition. As shown in Fig. \ref{fig:ndtidalall}, the left panel presents the electric tidal Love numbers $k_2^E$, and the right panel presents the magnetic tidal Love numbers $k_2^B$. The black dashed line in each panel represents the tidal Love numbers of the first excited state boson stars. It can be seen from both panels that for larger values of $\tilde{\omega}_0$ (specifically, $\tilde{\omega}_0=0.767, 0.766, 0.758$ in the figure), the curves develop spiral-like tails when the masses are small. In this region, $\tilde{\omega}_1$ is relatively large, approaching the turning point between the first and second branches. As $\tilde{\omega}_0$ increases, the degree of spiraling decreases until it eventually disappears. For the curves without such spiral-like tails, at fixed $\tilde{\omega}_0$, $k_2^E$ first decrease and then increase with increasing mass $M$, while $k_2^B$ increase monotonically. Moreover, the figure shows that the electric Love numbers are larger than the absolute values of the magnetic Love numbers. In both panels, the right endpoints of all solid lines intersect the black dashed line, indicating that the MSBSs degenerate into the first excited state boson stars.

 %%%%%%%%%%%%%%%%%%%%%%%%%%%%%%%%%%%%%%%%%%%%%%%%%%%%%%%%%%%		
 %%%%%%%%%%%%%%%%%%%%%%%%%%%%%%%%%%%%%%%%%%%%%%%%%%%%%%%%%%%

	%%%%%%%%%%%%%%%%%%%%%%%%%%%%%%%%%%%%%%%%%%%%%%%%%%%%%%%%%%
\section{CONCLUSION}\label{sec6}
	%%%%%%%%%%%%%%%%%%%%%%%%%%%%%%%%%%%%%%%%%%%%%%%%%%%%%%%%%%
    
In this paper, we studied the tidal Love numbers of MSBSs composed of ground state and first excited state complex scalar fields by first obtaining the background solutions. The properties of the field functions, ADM mass, and binding energy were discussed under synchronized and nonsynchronized frequency conditions. In both cases, the solutions can be classified into single-branch and double-branch types. For single-branch solutions, as $\tilde{\omega}$ or $\tilde{\omega}_1$ increases, the ground state field function increases, the first excited state field function decreases, and the ADM mass decreases. At the minimum and maximum frequencies, the MSBSs degenerate into the first excited state and the ground state boson stars, respectively. Double-branch solutions appear when $\tilde{\mu}_1 < 0.7976$ or $\tilde{\omega}_0 < 0.7675$. In each branch, the field functions and ADM mass vary with frequency in a way similar to the single-branch case. However, unlike the single-branch case, neither branch of the double-branch solutions degenerates into the ground state boson stars. As $\tilde{\mu}_1$ or $\tilde{\omega}_0$ decreases, the frequency range of the solutions shrinks. We used the binding energy to analyze the stability of MSBSs. Under the synchronized frequency condition, the double-branch solutions are all unstable. Stable solutions appear only in the single-branch case when $\tilde{\mu}_1 \ge 0.811$. In the nonsynchronized frequency case, the single-branch solutions always possess stable solutions. For the double-branch solutions, the first branch has stable solutions when $\tilde{\omega}_0$ is sufficiently large, while the second branch solutions are always unstable.

We calculated the quadrupolar ($\ell=2$) electric and magnetic tidal Love numbers of MSBSs by adding even-parity and odd-parity perturbations to the background. For stable MSBSs, we found that single-branch solutions under both synchronized and nonsynchronized frequency conditions exhibit a ``peak'' in the electric Love numbers, accompanied by a sign change, when the parameters satisfy $\tilde{\mu}_1 > 0.891$ or $\tilde{\omega}_0 > 0.777$. This behavior is related to the presence of the first excited state field and the closeness of the first excited state field mass to the ground state field mass. The magnetic tidal Love numbers show no such behavior and remain negative. For the first branch of the nonsynchronized frequency double-branch solutions, the electric Love numbers are all positive, while the magnetic tidal Love numbers are all negative, and the electric Love numbers are larger than the absolute values of the magnetic Love numbers. By comparison, we find that the Love numbers of MSBSs are larger in absolute value than those of the ground state boson stars. This means that the presence of the excited state makes the boson stars more easily deformable under the influence of an external tidal field.

Future work could study how rotation~\cite{Li:2019mlk} and self-interactions~\cite{Li:2020ffy} affect the tidal Love numbers of MSBSs. The model of multi-state Dirac stars has also been investigated, and the calculations presented in this paper can be similarly applied to multi-state Dirac stars~\cite{Liang:2023ywv}. Moreover, the calculation of tidal Love numbers can also be extended to other multi-field models~\cite{Ma:2023vfa,Ma:2023bhb,Liang:2022mjo}.

	%%%%%%%%%%%%%%%%%%%%%%%%%%%%%%%%%%%%%%%%%%%%%%%%%%%%%%%%%%
    \section*{ACKNOWLEDGMENTS}
	%%%%%%%%%%%%%%%%%%%%%%%%%%%%%%%%%%%%%%%%%%%%%%%%%%%%%%%%%%
	This work is supported by the National Natural Science Foundation of China (Grant No.~12275110 and No.~12247101) and the National Key Research and Development Program of China (Grant No.~2022YFC2204101 and No.~2020YFC2201503).

\end{document}